\documentclass[12pt,preprint]{aastex}

\begin{document}

\title{A Combined {\it Spitzer} and {\it Herschel} Infrared Study of Gas and Dust in the Circumbinary Disk Orbiting V4046 Sgr}

\author{Valerie A. Rapson\altaffilmark{1},
	Benjamin Sargent \altaffilmark{1},
	G. Germano Sacco\altaffilmark{2},
	Joel H. Kastner \altaffilmark{1},
	David Wilner\altaffilmark{3},
	Katherine Rosenfeld\altaffilmark{3},
	Sean Andrews\altaffilmark{3},
	Gregory Herczeg\altaffilmark{4},
	Nienke van der Marel\altaffilmark{5}
		}
\email{var5998@rit.edu}

\altaffiltext{1}{School of Physics and Astronomy, Rochester Institute of Technology, 1 Lomb Memorial Drive, Rochester, NY 14623-5603, USA}
\altaffiltext{2}{INAF-Arcetri Astrophysical Observatory, Largo Enrico Fermi 5, I - 50125 Florence, Italy}
\altaffiltext{3}{Harvard-Smithsonian Center for Astrophysics, 60 Garden Street, Cambridge, MA 02138, USA}
\altaffiltext{4}{Kavli Institute for Astronomy and Astrophysics, Peking University, Yi He Yuan Lu 5, Haidian Qu, Beijing 100871, China}
\altaffiltext{5}{Leiden Observatory, Leiden University, PO Box 9513, 2300 RA, Leiden, The Netherlands}

\begin{abstract}
We present results from a spectroscopic {\it Spitzer} and {\it Herschel} mid-to-far-infrared study of the circumbinary disk orbiting the evolved (age $\sim$12-23 Myr) close binary T Tauri system V4046 Sgr. {\it Spitzer} IRS spectra show emission lines of [Ne~{\sc ii}], H$_{2}$ S(1), CO$_{2}$ and HCN, while  {\it Herschel} PACS and SPIRE spectra reveal emission from [O~{\sc i}], OH, and tentative detections of H$_{2}$O and high-J transitions of CO. We measure [Ne~{\sc iii}]/[Ne~{\sc ii}] $\lesssim$ 0.13, which is comparable to other X-ray/EUV luminous T Tauri stars that lack jets. We use the H$_{2}$ S(1) line luminosity to estimate the gas mass in the relatively warm surface layers of the inner disk. The presence of [O~{\sc i}] emission suggests that CO, H$_2$O, and/or OH is being photodissociated, and the lack of [C {\sc i}] emission suggests any excess C may be locked up in HCN, CN and other organic molecules. Modeling of silicate dust grain emission features in the mid-infrared indicates that the inner disk is composed mainly of large (r$\sim$ 5 $\mu$m) amorphous pyroxene and olivine grains ($\sim$86\% by mass) with a relatively large proportion of crystalline silicates. These results are consistent with other lines of evidence indicating that planet building is ongoing in regions of the disk within $\sim$30 AU of the central, close binary. 
\end{abstract}

\section{Introduction}
The relatively rare examples of nearby T Tauri stars (i.e., those within $\sim$100 pc of Earth) are excellent candidates for studying planet formation processes. Their proximity allows us to analyze the contents and structure of circumstellar disks in which young protoplanets are likely forming. Near to far-infrared spectroscopy allows us to probe the contents of these protoplanetary disks, as the strongest emission from such disks arises at infrared and (sub)-mm wavelengths. Mid-IR (5-40 $\mu$m) spectroscopy traces both the gas and dust components of the inner (R $<$10 AU) planet forming regions of the disk. At medium to high resolution, we detect emission features due to increasing disk temperature from the disk midplanes to their exteriors \citep{calvet1991, malbet1991}. In most cases, such spectral
features only probe the disk atmosphere, as the disk becomes optically thick towards the midplane. Emission features may also arise from optically thin
regions of the disk due to dust sublimation or grain growth and planetesimal formation \citep{pontoppidan2014}. Transition disks with an inner gap
carved out by giant planet formation and/or disk photoevaporation may display cavity walls with emission features due to being directly exposed to stellar UV radiation \citep[e.g.][]{cleeves2011}. Far-IR (50-700 $\mu$m) spectroscopy traces emission from colder gas and dust in the outer (R $\gtrsim$10 AU) disk. At these longer wavelengths, we expect emission features from gas mass tracers such as CO and atomic species that trace photodissociation regions at disk surfaces.

Previous {\it Spitzer} space telescope and ground based mid-IR spectroscopic studies of pre-main sequence (pre-MS) star/disk systems have revealed a variety of atomic and molecular emission features \citep[e.g.][]{salyk2011b, pontoppidan2010}. Molecules such as C$_{2}$H, HCN, CN, OH, H$_{2}$O and C$_{2}$H$_{2}$, whose abundances are sensitive to stellar  UV and X-ray radiation, are used to gauge the strength of disk gas photodissociation and radiative heating processes \citep[e.g.][]{pascucci2013, pascucci2009, henning2010, bergin2003}. H$_{2}$O has also been observed in the planet forming regions around many young stars \citep[e.g.][]{salyk2011a}. Metal ions such as [Ne~{\sc ii}] have also been detected towards transitional disks \citep{sacco2012,pascucci2007,lahuis2007,gudel2010} and may trace the influence of high energy radiation on disk photoevaporation rates. 

Far-IR spectroscopic observations of pre-MS star/disk systems via the {\it Herschel} space observatory help constrain the disk gas and dust mass, as well as the effects of high energy radiation on outer disk heating and chemistry \citep[e.g.][]{fedele2013, meeus2013, riviere_marichalar2013, howard2012, thi2010} . Far-IR emission from pure rotational CO lines probe temperature regimes between those of the near-IR vibrational CO lines and the low-J sub-mm/radio CO lines and can reveal information about this distribution of gas throughout the disk \citep[e.g.][]{bruderer2012}. The presence of far-IR [O~{\sc i}], [C {\sc i}] and [C {\sc ii}] emission lines may indicate that UV or X-ray radiation is incident on the outer disk and photodissociating CO or OH molecules \citep{mamon1988,aresu2012}. Modeling these and other diagnostic emission lines in {\it Herschel} spectra, combined with models of {\it Spitzer} spectroscopic observations, can constrain total disk mass, as well as the composition and distribution of gas and dust throughout the disk.

Here, we present a combined spectroscopic {\it Spitzer} and  {\it Herschel} study of the protoplanetary disk orbiting the nearby \citep[D $\sim$73 pc;][]{torres2008} pre-MS binary V4046 Sgr. V4046 Sgr is an isolated, spectroscopic binary that is a member of the $\beta$ Pic Moving Group, with an age $\sim$12-23 Myr old \citep{torres2006, binks2013, mamajek2014}. The binary consists of nearly equal mass components with spectral types K5Ve and K7Ve \citep{stempels_gahm2004} and total mass 1.75 M$_{\odot}$ \citep{rosenfeld2012b}. The stars are separated by 0.04 AU \citep[p $\sim$ 2.4 days;][]{stempels_gahm2004} and are surrounded by a large (R $\sim$370 AU) circumbinary disk of gas and dust inclined at $\sim33^{\circ}$ from face on \citep{rosenfeld2012b, rodriguez2010}. The central stars are actively accreting material \citep{stempels_gahm2004, donati2011, argiroffi2012} from the disk, which has an inner gap of radius $\sim30$ AU that is possibly due to ongoing Jovian planet formation \citep{rosenfeld2013}. The disk is also evidently undergoing photoevaporation by high energy radiation from the central stars \citep{sacco2012}, which, in conjunction with planet formation, may contribute to the inner disk clearing \citep[][and references therein]{Alexander2014}

In this paper, we present a census of atomic and molecular emission features detected in the Spitzer and Herschel spectra of V4046 Sgr, and we model the continuum of the Spitzer spectrum to determine the dust particle composition of the disk. Our {\it Spitzer} and {\it Herschel} observations and data reduction processes are discussed in Section 2 and in Sections 3 \& 4 we report the results of our emission line study. In Section 5 we discuss our modeling of the dust features in the {\it Spitzer} spectrum and we present a summary of our conclusions in Section 6. 

\section{Observations and Data Reduction}

\subsection{{\it Spitzer} IRS}
Low (R $\sim$60-130) and high (R $\sim$600) resolution {\it Spitzer} Space Telescope \citep{werner2004} data for V4046 Sgr were obtained with the InfraRed Spectrograph \cite[IRS;][]{houck2004} in April 2005 (PI: Mitsuhiko Honda\footnote{program ID: 3580, AORKEY: 11197440}), and were retrieved through the {\it Spitzer} Heritage Archive.  Short low (SL, 5-20$\mu$m), short high (SH, 10-20 $\mu$m), and long high (LH,  20-37 $\mu$m) data were reduced individually using SMART v8.2.5 \citep{higdon2004}. No background subtraction or low-level rogue pixel removal was performed on the high resolution data because no dedicated background images were available. All images were manually examined after processing and bad pixels (particularly near 6.9 and 17.6 $\mu$m) were smoothed over. 

The two nods of each set of cleaned spectra were averaged together and the three modes were combined using
custom IDL programs. SL data are available for the $\sim$5-14 $\mu$m range, but are used only in the 5-10 $\mu$m range, where high resolution data are not available. Examination of the original spectra showed a relatively smooth continuum, with a small jump between the SH and LH spectra at $\sim$20 $\mu$m. As V4046 Sgr is unresolved by {\it Spitzer}, this mismatch is likely due to the source falling off the slit in the SH module. Hence, the SL and SH spectra were scaled by a multiplicative factor of 1.182 such that the SL and SH matched the LH spectra at $\sim$20 $\mu$m. The final resulting {\it Spitzer} spectrum (5-35 $\mu$m) is shown in Figure \ref{spitzer_spec}. It is possible that the excess flux in the LH spectrum is due to residual sky background. If this is the case, then the LH spectra should be scaled down by 1.182 to match that of the SH and SL spectra.

\subsection{{\it Herschel} PACS and SPIRE}

\begin{figure}
\includegraphics[scale=0.8]{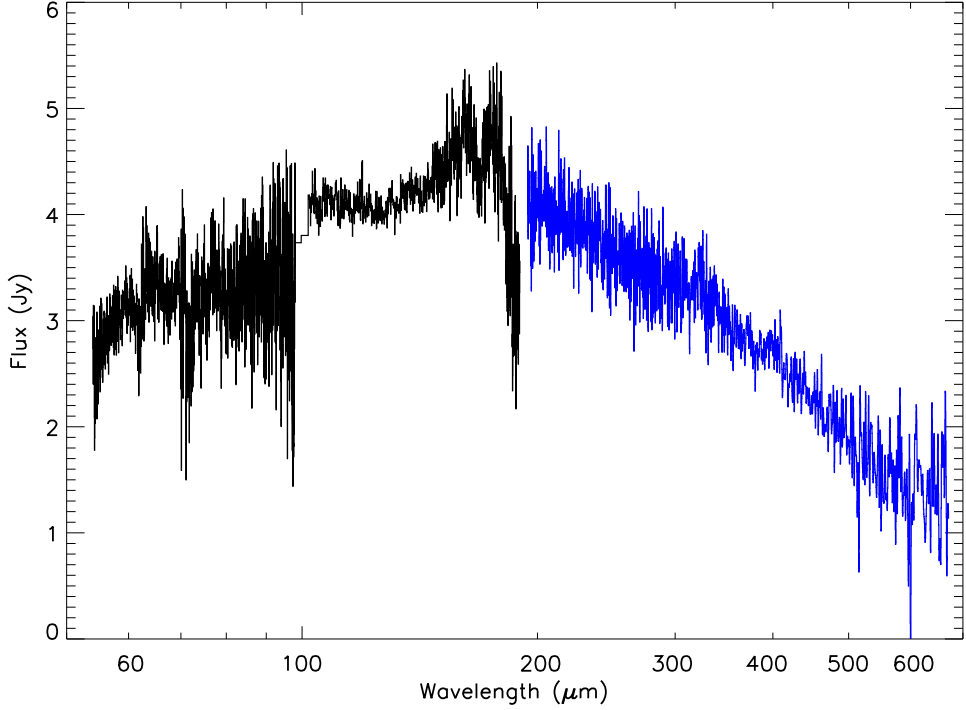}
\caption{\label{herschel_spec} Top: {\it Herschel} PACS (black) and SPIRE (blue) spectra of V4046 Sgr. The data have been rebinned by a factor of 3. }
\end{figure}

{\it Herschel} Space Observatory \citep{pilbratt2010} Photodetector Array Camera and Spectrometer \citep[PACS; ][]{poglitsch2010} range scan spectra and Spectral and Photometric Imaging Receiver \citep[SPIRE;][]{griffin2010} spectra of V4046 Sgr were obtained as part of the Cycle 1 General Observer Program\footnote{Observation IDs: 1342231043, 1342242448, 1342242449} (PI: G. Sacco) in June and July of 2012. SPIRE data were obtained at medium resolution (R $\sim$160), and PACS data were obtained in SED mode (R $\sim$940-5500). Together, these observations cover the 55-670 $\mu$m range. Examination of the SPIRE level 2 data products showed a discontinuity between the two bands (190-310 $\mu$m and 300-670 $\mu$m), suggesting that the background was incorrectly subtracted by the pipeline\footnote{See section 6.4 in the SPIRE data reduction Guide v2.1.}. We therefore reduced both the PACS and SPIRE data from level 0 in HIPE v10 \citep{ott2010} using interactive background normalization scripts, which allow us to subtract the background more accurately than is done in the standard pipeline. Data from calibration trees 48 and 10.1 for PACS and SPIRE, respectively, were used to correct for instrumentation effects and to flux calibrate the spectra. The center spatial pixel (spaxel) was extracted from the resulting 5x5 spaxel array and order edges and regions in the spectrum where light leakage occurs were removed. The resulting {\it Herschel} PACS and SPIRE data are shown in Figures \ref{herschel_spec} and \ref{herschel_zoom}.

\begin{figure}
\includegraphics[scale=0.8]{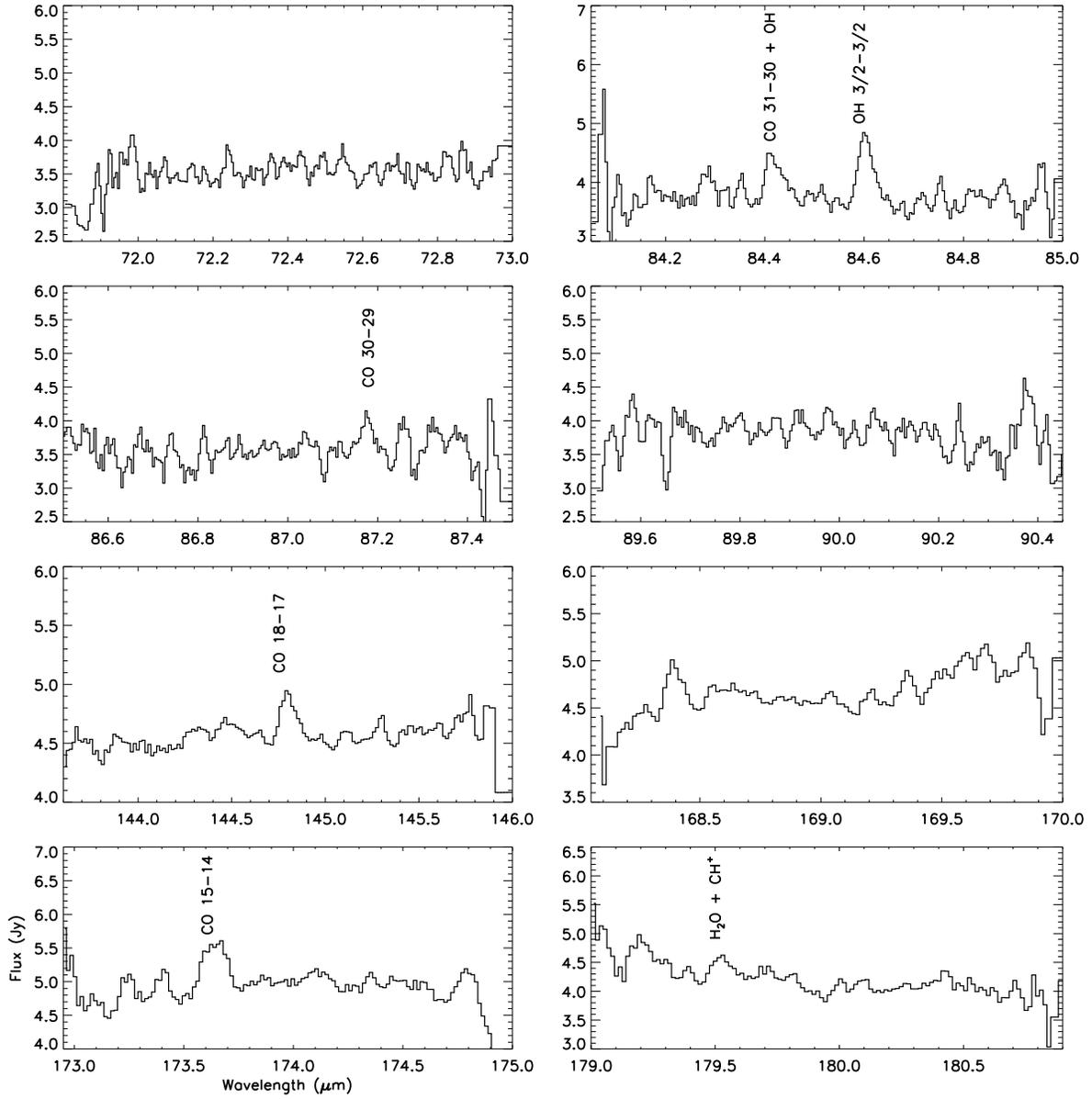}
\caption{\label{herschel_line} {\it Herschel} PACS line scan spectroscopy data with emission features labeled.}
\end{figure}

{\it Herschel} PACS line scan spectra were obtained in Cycle 2\footnote{Observation IDs: 1342269454 and 1342269455} (P.I. G. Herczeg) in June 2013 with the goal of searching for mid- to high-J lines of CO, as well as H$_{2}$O and OH. These data cover eight different wavelength regions. The PACS line scan data were reduced from level 0 using background normalization scripts in HIPE v11 with PACS calibration tree 56. Two output spectra were produced, one that was flux calibrated using the central 3x3 spaxel region and one that was not. Telescope jitter and slight pointing errors can cause some of the source flux in {\it Herschel} observations to extend beyond the central spaxel. This effect can be corrected by taking the flux ratio of the sum of the central 3x3 spaxels to the single center spaxel and comparing it with the ratio of a perfectly pointed observation of a point source, then adjusting the source flux accordingly.  A comparison between the 3x3 corrected spectra and the non-corrected spectra shows a $\sim$10\% difference in continuum flux values. Since this difference is small, and since the PACS SED spectra were not 3x3 corrected, we chose to use the non-3x3 corrected spectra. The continuum flux of the PACS line scan and SED data for each region also matched to within $\sim$10\%, consistent with the level of systematic error. The PACS line scan spectra are shown in Figure \ref{herschel_line}.

\section{Analysis: Gas Emission Features}

\subsection{Emission line identification}

Spectral line identification in both the {\it Spitzer} and {\it Herschel} spectra was
performed both by eye and using custom IDL programs. To identify potential lines in the IRS
spectrum, we first fit the continuum regions of the SH and LH data with a high
order polynomial, then subtracted the data from the fitted continuum. We then
identified candidate emission lines as features that span at least two spectral bins and are at least 2$\sigma$ above the continuum (where $\sigma$ is defined as the standard deviation of the continuum subtracted spectra). To confirm which of these detections are likely emission lines and not spurious detections or dust features, we subtract the continuum dust model (see section 3.1) from the original spectra, and repeat the above line detection process. The NIST atomic spectral database\footnote{$http://physics.nist.gov/PhysRefData/ASD/lines\_form.html$}, the Splatalogue database for astronomical spectroscopy\footnote{$http://splatalogue.net/$}, and previous publications pertaining to mid- to far-infrared emission features from protoplanetary disks were used to determine the species and transition potentially associated with each candidate emission feature. We restrict these identifications to spectral features that are at least 2$\sigma$ above the continuum and that correspond to molecular or atomic features previously identified in protoplanetary disks. Table \ref{lines} lists the candidate lines identified in the {\it Spitzer} spectrum and their measured wavelengths, energies, fluxes, full widths at half maximum, and equivalent widths. 

Emission features in the {\it Herschel} PACS and SPIRE spectra were determined in similar fashion. The {\it Herschel} PACS range scan and SPIRE spectra are less sensitive than the {\it Spitzer} and PACS line spectra, and thus only a few candidate emission features are identified at levels $>$ 3$\sigma$. Table \ref{lines_herschel} lists the candidate lines identified in all {\it Herschel} spectra and their measured wavelengths, energies, fluxes, full widths at half maximum, and equivalent widths.

\subsubsection{{\it Spitzer} IRS molecular \& atomic line inventory}

\begin{figure}
\includegraphics[scale=0.8]{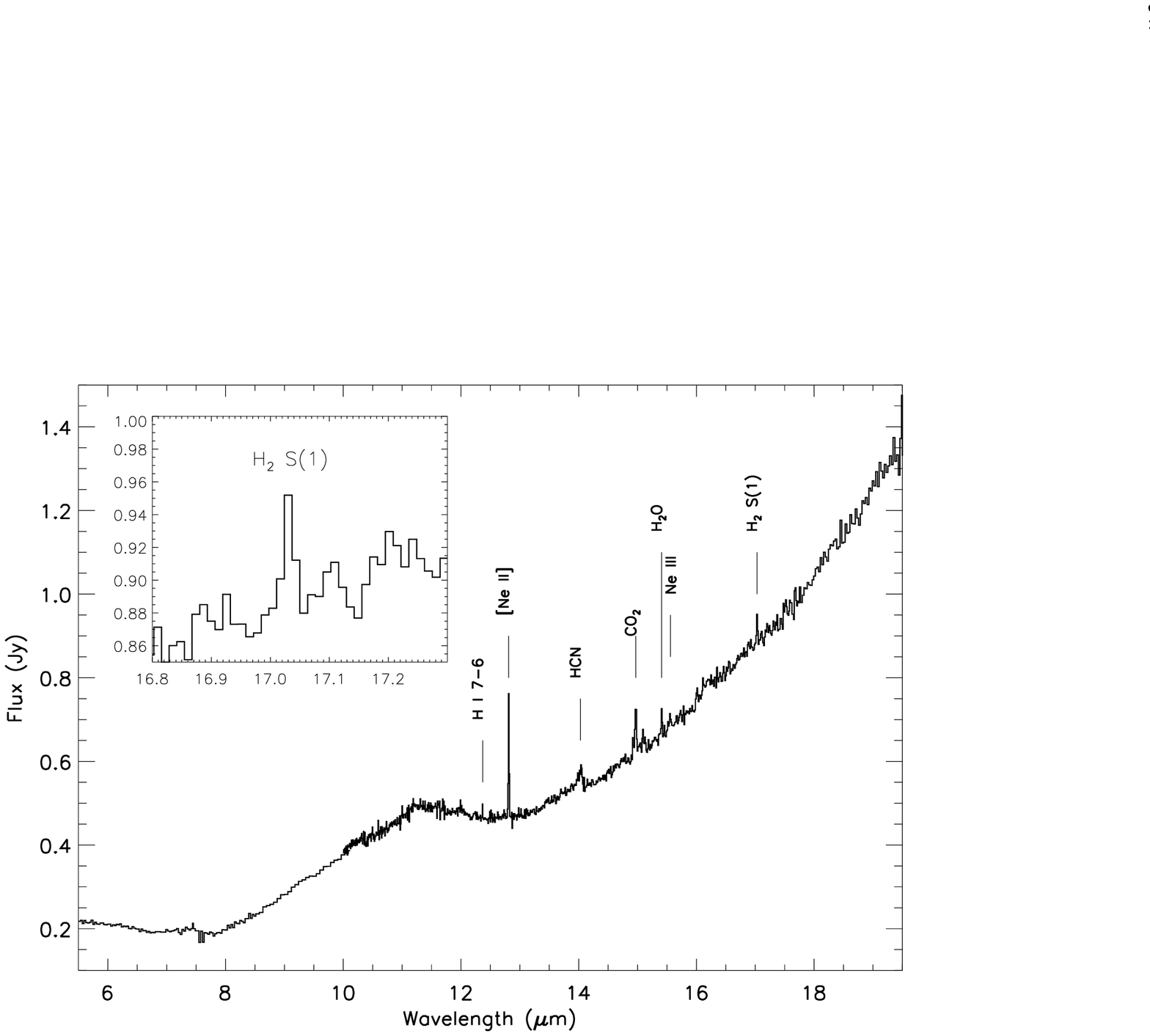}
\includegraphics[scale=0.8]{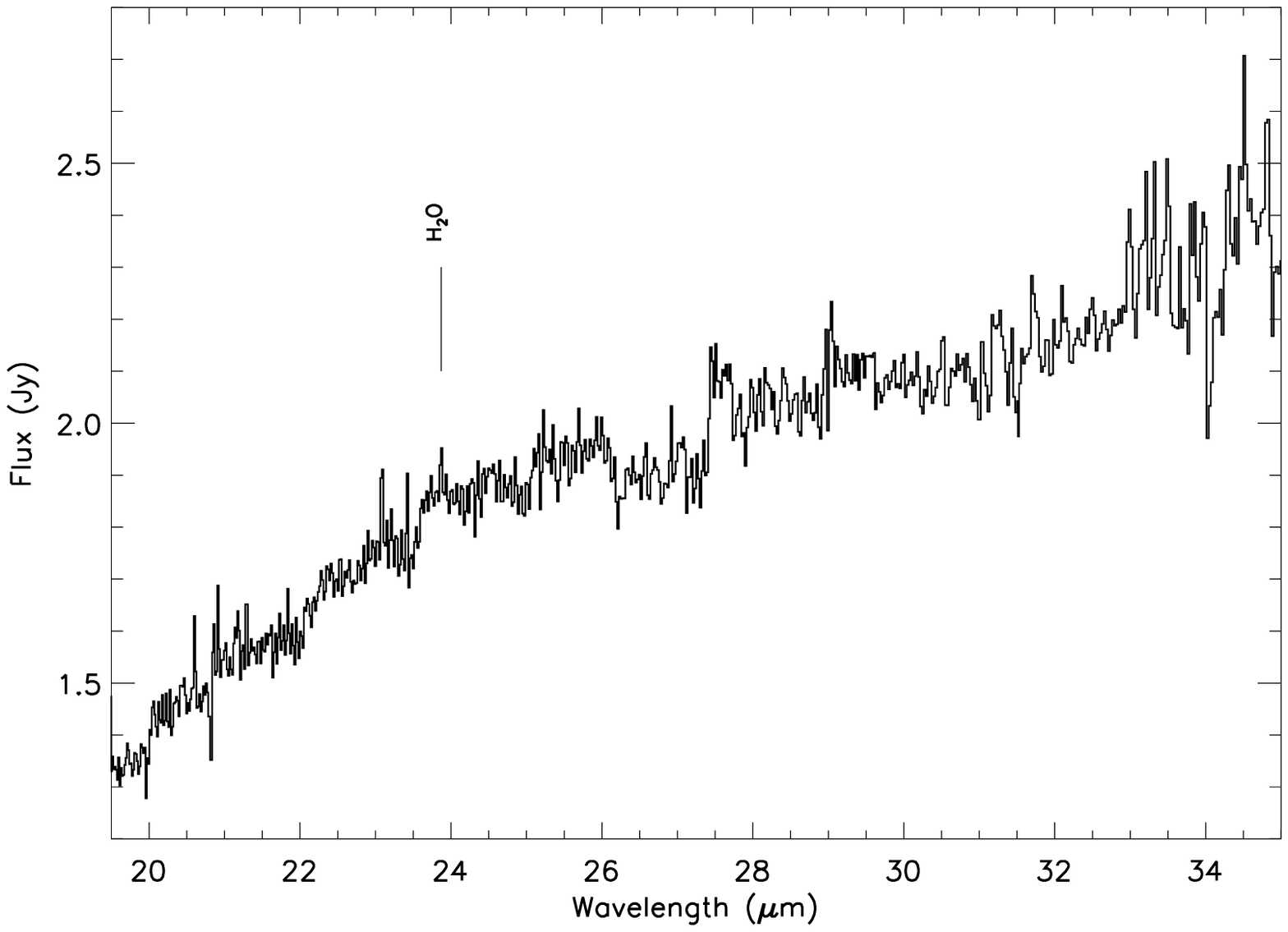}
\caption{\label{spitzer_spec} {\it Spitzer} IRS spectrum of V4046 Sgr with emission features labeled. Top: 5.5-19.5 $\mu$m spectrum with an inset showing the H$_{2}$ 0-0 S(1) (J=3$\rightarrow$1) line at 17.035 $\mu$m. Bottom: 19.5-35 $\mu$m spectrum with emission features labeled.}
\end{figure}

\begin{figure} 
\includegraphics[scale=0.5]{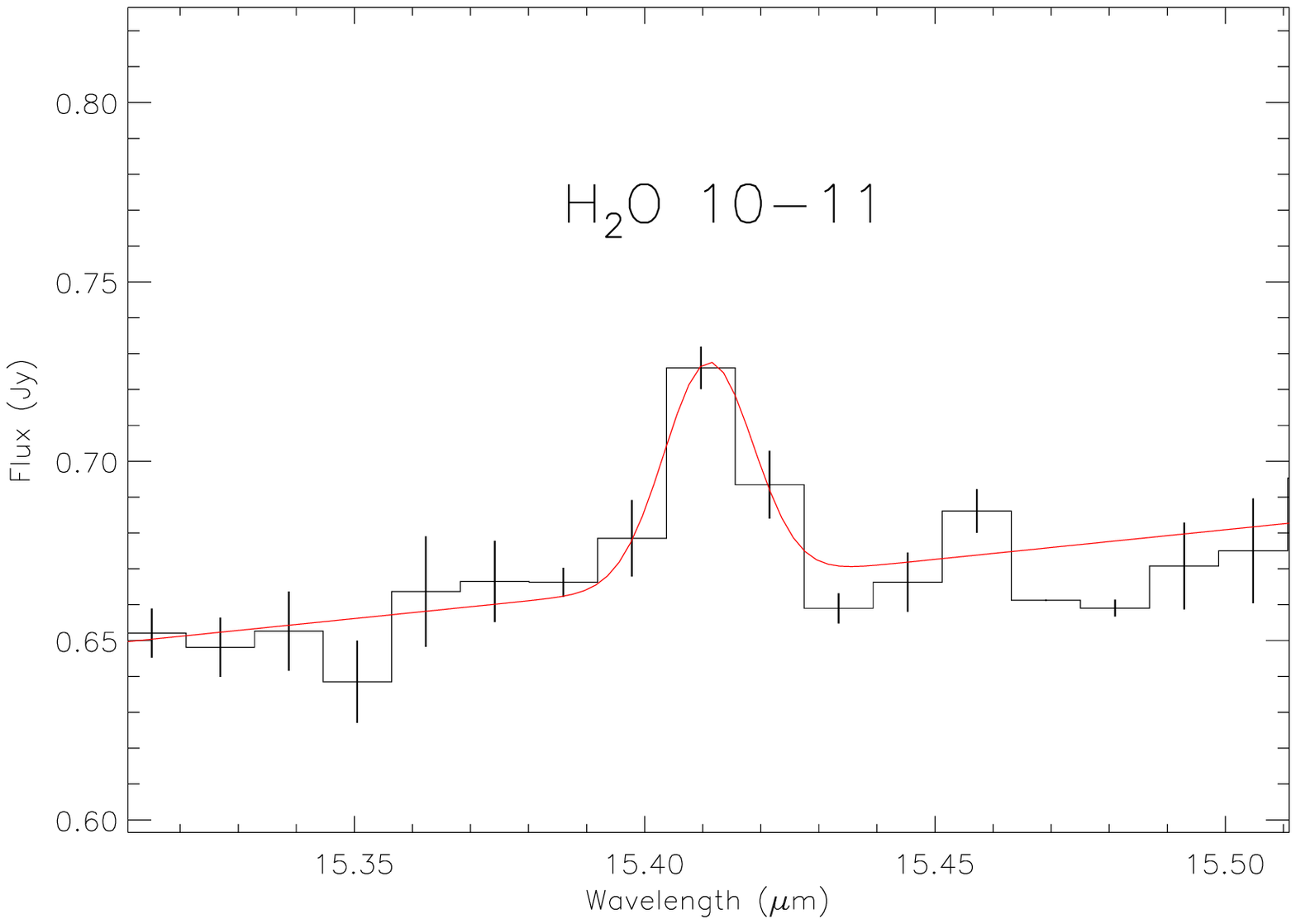}
\includegraphics[scale=0.5]{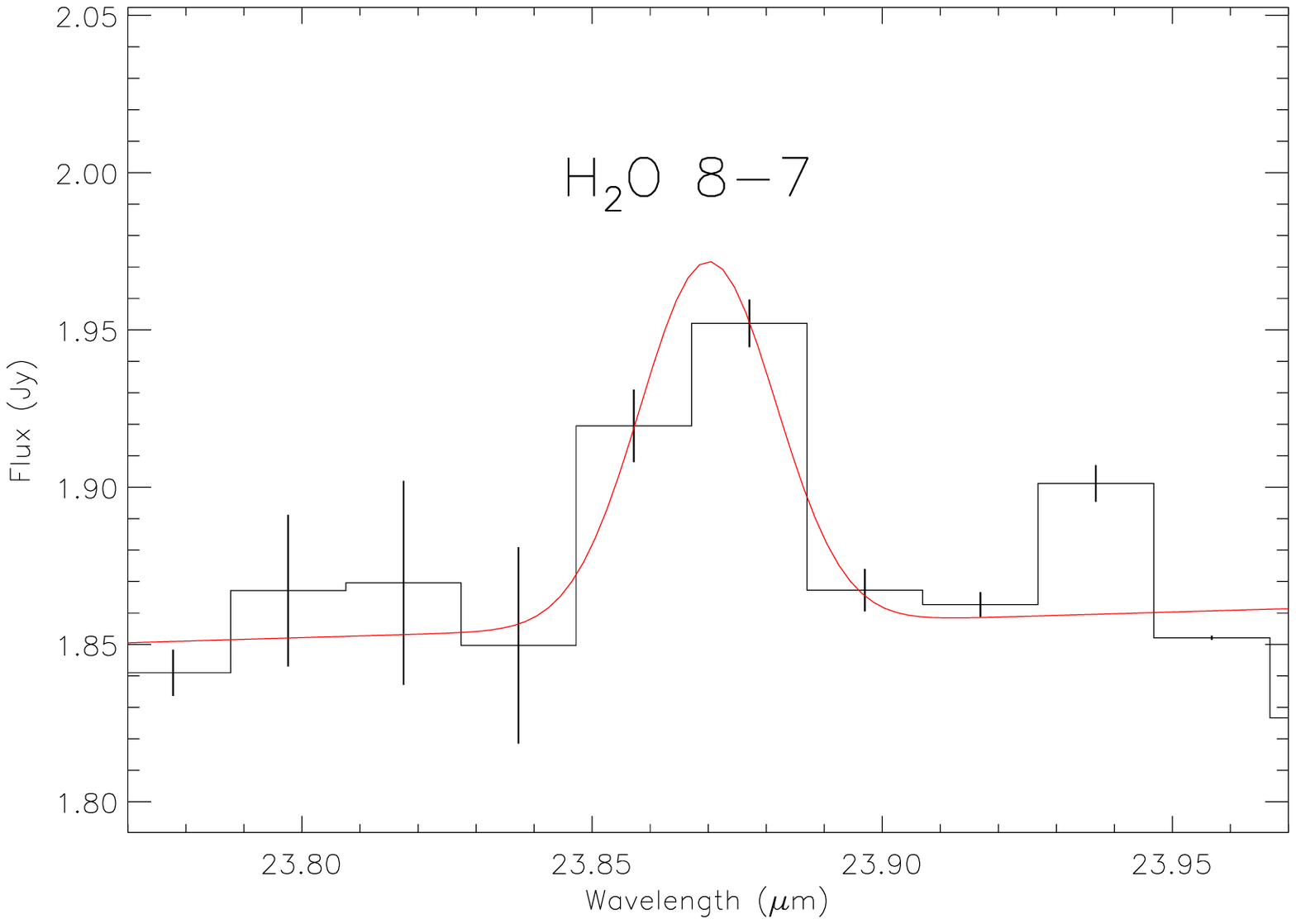}
\caption{\label{spitzer_spec_close} H$_2$O emission features tentatively detected in the {\it Spitzer} IRS spectrum of V4046 Sgr with Gaussian fits overlaid}
\end{figure}

The disk around V4046 Sgr shows evidence for many different atomic and molecular species (Figure \ref{spitzer_spec}, \ref{spitzer_spec_close} and Table \ref{lines}). We detect strong emission from [Ne~{\sc ii}], as well as emission lines of H$_{2}$ S(1), CO$_{2}$ and HCN in the {\it Spitzer} spectrum.  Emission lines of these species are commonly seen  in T Tauri star/disk systems \citep[e.g.][]{pascucci2013,salyk2011a, pontoppidan2010,  pascucci2009, salyk2008, najita2008}. The {\it Spitzer} spectrum also shows H {\sc i} 7-6 emission at 12.4 $\mu$m, which likely does not originate within the disk but, rather, arises from the accretion shock regions at the (binary) central stars and/or from their stellar corona \citep{pascucci2007,rigliaco2015}.

Emission lines from H$_{2}$O at 15.2 and 17.2 $\mu$m, which are often used as markers to establish that  a disk is water-rich \citep[e.g.][]{pontoppidan2010}, are not evident in the Spitzer spectrum of V4046 Sgr. The spectrum may display H$_{2}$O features at  15.4 and 23.9 $\mu$m, as well as possible complexes of H$_{2}$O features near 25 and 31 $\mu$m (see Figure \ref{spitzer_spec}). The detailed, self-consistent modeling of H$_2$O and OH emission required to confirm these identifications will be pursued in followup work. V4046 Sgr also shows HCN emission at 14.03 $\mu$m, but lacks the C$_{2}$H$_{2}$ emission lines typically found in younger T Tauri systems \citep{salyk2011a, carr2008}. \citet{pontoppidan2010} and \citet{salyk2011a} find that H$_{2}$O, OH and other molecular gas emission lines originating from warm gas in the inner disk are weaker in most transitional disk systems, likely because these disks are beginning to form an inner disk clearing either due to planet formation or disk photoevaporation. Therefore, the lack of strong H$_{2}$O and OH emission lines, and the generally weak levels of molecular emission in V4046 Sgr, is consistent with the fact that this system is a transitional disk with an inner clearing \citep{rosenfeld2013}. 

\subsubsection{{\it Herschel} PACS+SPIRE molecular \& atomic line inventory}

\begin{figure}
\includegraphics[scale=0.5]{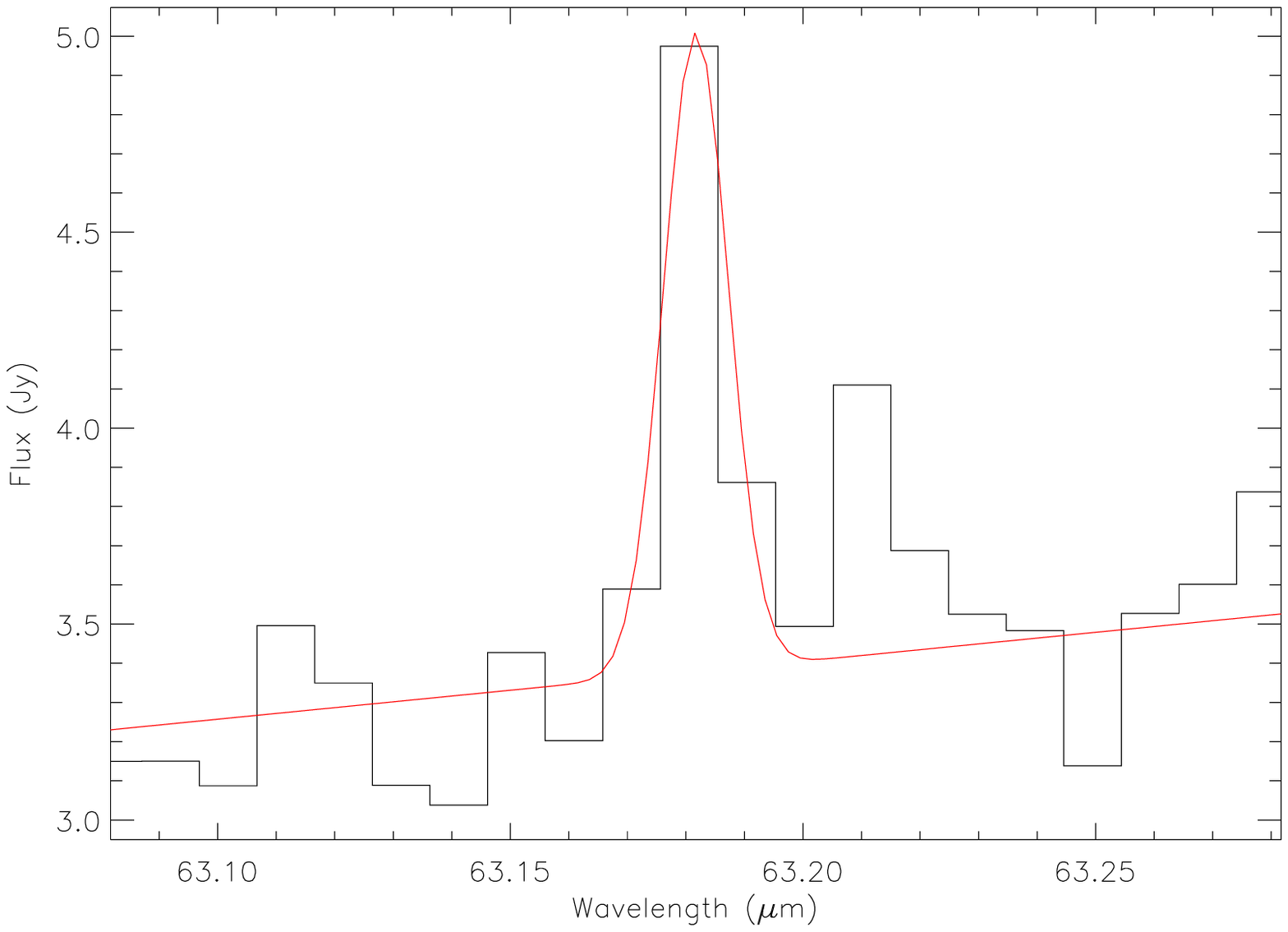}
\includegraphics[scale=0.5]{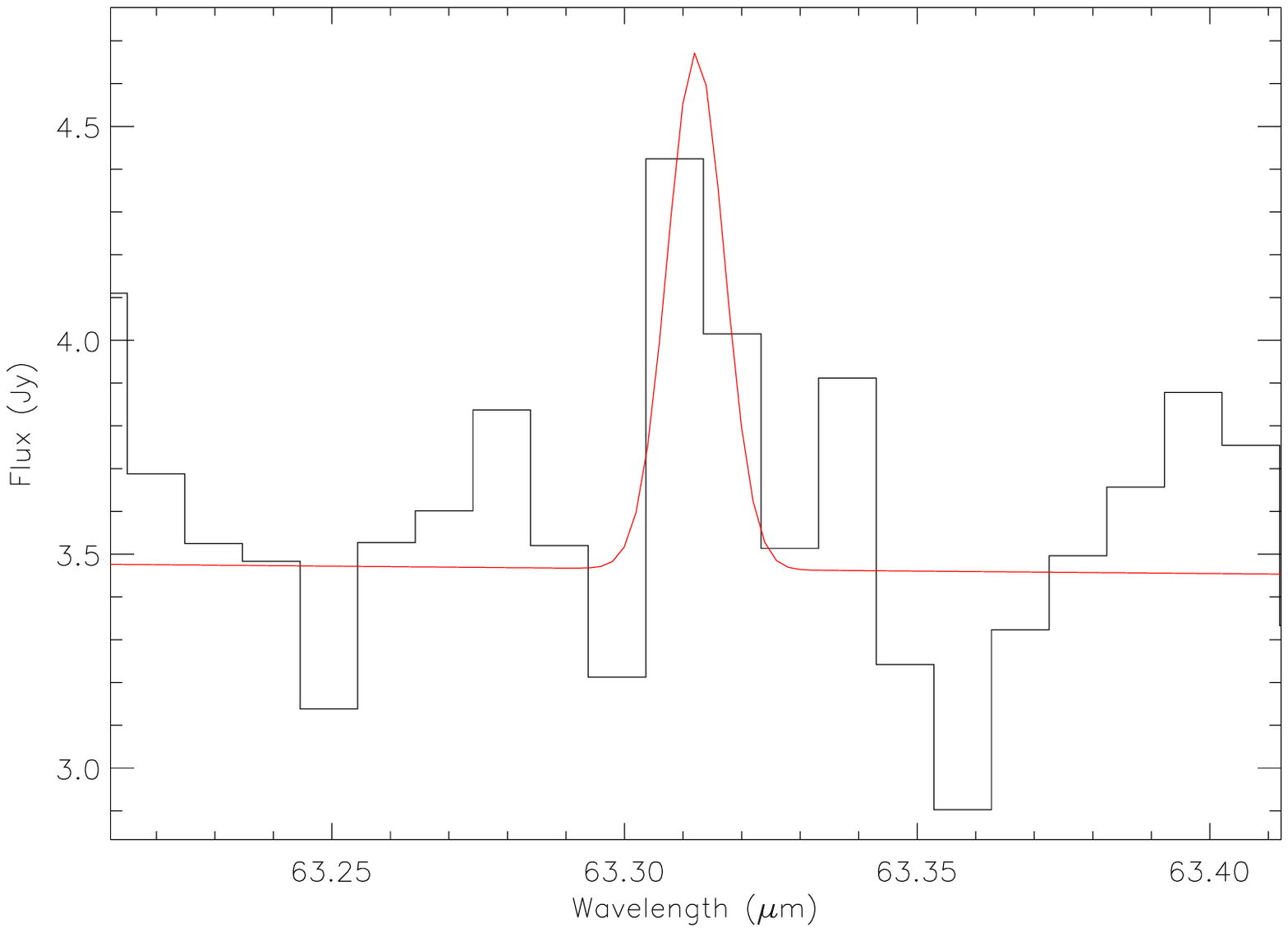}
\includegraphics[scale=0.5]{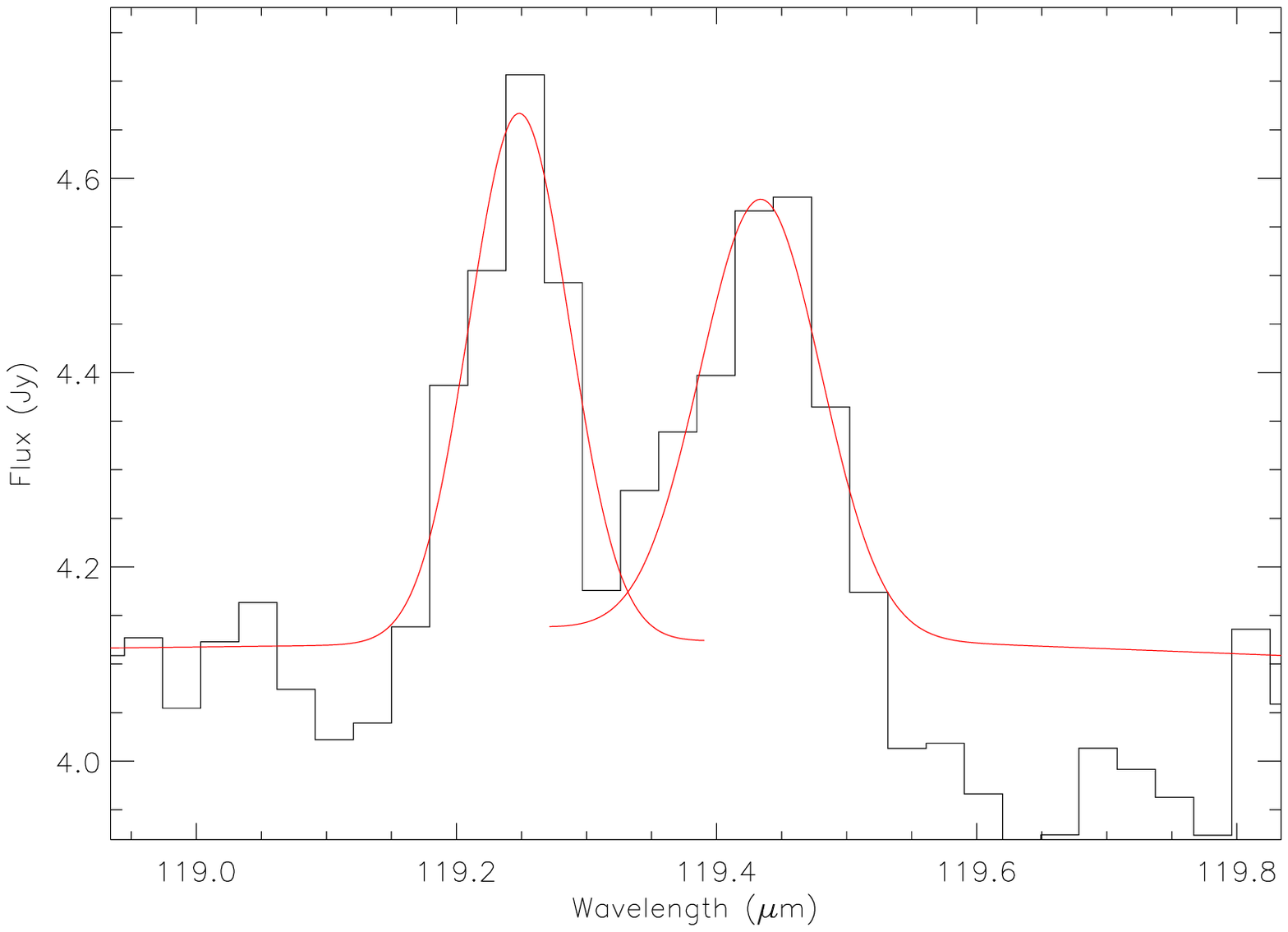}
\includegraphics[scale=0.5]{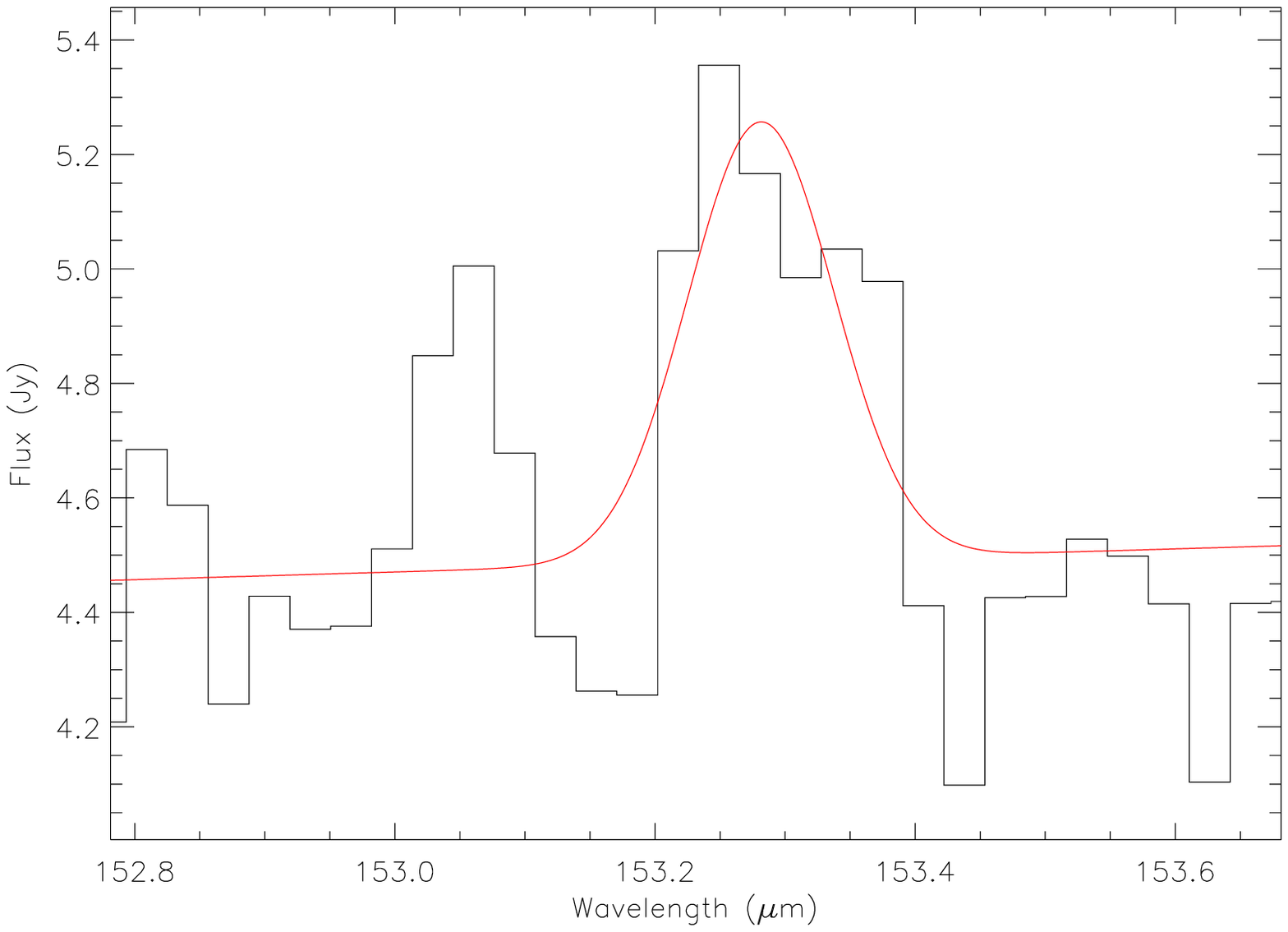}
\caption{\label{herschel_zoom} [O {\sc i}]  (top left), o-H$_{2}$O (top right), OH $^{2}\Pi_{3/2}$ J= $\frac{3}{2}\rightarrow\frac{5}{2}$ (bottom left) and CO J=17$\rightarrow$16  (bottom right) emission features detected (or tentatively detected) in the unbinned V4046 Sgr Herschel PACS SED spectrum with Gaussian fits overlaid (red). }
\end{figure}

Emission features in the {\it Herschel} PACS SED and SPIRE spectra and line IDs are presented in Figures \ref{herschel_spec} and \ref{herschel_zoom} and Table \ref{lines_herschel}. The {\it Herschel} spectra were obtained with the goal of searching for emission from [O~{\sc i}] at 63 $\mu$m and [C {\sc i}] at 370 $\mu$m arising from UV photodissociation of CO molecules in the outer disk \citep{mamon1988}. We report the detection of [O~{\sc i}] emission (Table \ref{lines_herschel}, Figure \ref{herschel_zoom}), but did not detect [C {\sc i}]. We also detect emission from an OH 3/2-3/2 doublet at 119.2 and 119.4 $\mu$m (Figure \ref{herschel_zoom}), which likely arises from UV photodissociation of H$_{2}$O in the outer disk. One high level CO ro-vibrational transition (J=17 $\rightarrow$ 16) is tentatively detected and 3$\sigma$ upper limits for all other ro-vibrational transitions of CO between 65 $\mu$m and 600 $\mu$m range from 0.42 -- 4.68 $10^{-14}$ erg s$^{-1}$ cm$^{-2}$.

The {\it Herschel} PACS line scan spectra cover multiple limited spectral regions also spanned by PACS range spectroscopy, but achieved higher sensitivity and spectral resolution. These observations reveal several emission features that were below the detection limit of the PACS SED data (Figure \ref{herschel_line} and Table \ref{lines_herschel}). Specifically, we detect (or tentatively detect) emission from four high-level CO ro-vibrational transitions (J= 31 $\rightarrow$ 30, 30 $\rightarrow$ 29, 18 $\rightarrow$ 17, and 15 $\rightarrow$14), as well as OH and CH$^{+}$. The PACS line scan spectra do not cover the spectral range in which emission features were detected in the PACS SED spectra, so we cannot use these data to confirm the existence of the H$_2$O and CO J=17 $\rightarrow$ 16 features.

\section{Discussion: Gas Emission Features}

V4046 Sgr exhibits emission features that allow us to further understand the structure, content, and chemistry of the circumbinary disk. We can use these emission features to investigate disk photoevaporation and photodissociation processes, and to compare the V4046Sgr system with other similarly evolved systems such as TW Hya \citep[age $\sim$8 Myr, d $\sim$ 53 pc;][and references therein] {ducourant2013}, and other T Tauri star-disk systems in nearby associations. Below, we comment on notable features in the {\it Spitzer} and {\it Herschel} spectra of V4046 Sgr and combine the {\it Spitzer} and {\it Herschel} spectra to obtain a full mid to far-IR SED (5-670 $\mu$m) so as to compare the SED with a dust disk model from \citet{rosenfeld2013}.

\subsection{H$_{2}$}

Although H$_{2}$ is the most abundant species in circumstellar disks, the electric quadrupole nature of its rotational transitions result in weak emission that is difficult to detect. However, pure rotational H$_{2}$ emission has been detected in the mid-IR around some T Tauri and Herbig Ae/Be stars \citep[i.e.][]{carr2011,najita2010, bitner2008, carmona2008, bary2008, lahuis2007}, and even in a few disks around young brown dwarfs \citep{pascucci2013}. This emission may be due to collisional excitation, X-ray/UV irradiation of the disk surface, and/or accretion shocks onto the disk \citep[and references therein]{bitner2008}. 

We detect emission from pure rotational H$_{2}$ 0-0 S(1) (J=3$\rightarrow$1) at 17.035 $\mu$m in the {\it Spitzer} spectrum (Table \ref{lines}). Since V4046 Sgr lies far from any molecular clouds  \citep{kastner2008}, the H$_2$ emission most likely arises from warm gas in the surface layers of the circumbinary disk. The line flux of H$_{2}$ is $\sim$1.50$\times$10$^{-14}$ erg s$^{-1}$ cm$^{-2}$ (L$_{H_2}=$1.04$\times$10$^{28}$ erg s$^{-1}$). This is comparable to the 17 $\mu$m H$_{2}$ line strengths measured in T Tauri stars studied by \citet{bitner2008}, as well as to the 17$\mu$m H$_{2}$ line luminosity from the (similarly isolated) TW Hya \citep[L$_{H_2}=$4.17$\times$10$^{27}$ erg s$^{-1}$, assuming a distance of 53 pc;][]{najita2010}.

 \citet{carmona2008} did not detect H$_{2}$ emission towards V4046 Sgr using VISIR on the VLT Melipal telescope. They report a three sigma upper limit for 17 $\mu$m H$_{2}$ emission of $<$1.3$\times$10$^{-14}$ erg s$^{-1}$ cm$^{2}$, which is roughly equivalent to our measured H$_{2}$ 17 $\mu$m flux. The non-detection of H$_{2}$ by \citet{carmona2008} may be due to VISIR having a smaller slit size, or the difficulty of removing atmospheric effects from ground-based spectra. It is also possible that the H$_{2}$ emission is variable as a result of ongoing accretion and variable X-ray/UV luminosity \citep{bary2008}.

Assuming that the H$_{2}$ emission is optically thin, the H$_{2}$ gas is in local thermal equilibrium, and the IRS slit covers the entire source (note that we have already corrected for the source possibly extending beyond the coverage of the SH slit), we can estimate an upper limit for the mass of the emitting H$_{2}$ gas as a function temperature from
\begin{equation}
M_{gas}=f\times1.76\times10^{-20}\frac{4\pi d^{2}\lambda F_{ul}}{hcA_{ul}x_{u}(T)}\; M_{\odot}, 
\end{equation}
where $F_{ul}$ is the flux of the H$_{2}$ emission line, $\lambda$ is the central wavelength of the emission line, $d$ is the distance in parsecs, A$_{ul}$ is the Einstein coefficient of the J $=u-l$ transition\footnote{A$_{ul}$=4.76$\times$10$^{-10}$ s$^{-1}$ for the H$_{2}$-ortho J = 3-1 transition \citep{Wolniewicz1998}}, and $x_{u}(T)$ is the population at level $u$ for a given excitation temperature $T$  \citep{thi2001}. Since only the H$_{2}$-ortho 17 $\mu$m emission line is detected in our spectrum, we multiply by a conversion factor $f= 1+(ortho/para)^{-1}$ \citep{carmona2008}, where the equilibrium ortho/para ratio can be determined from equation (1) in \citet{takahashi2001}. 

\begin{figure}
\includegraphics[scale=1]{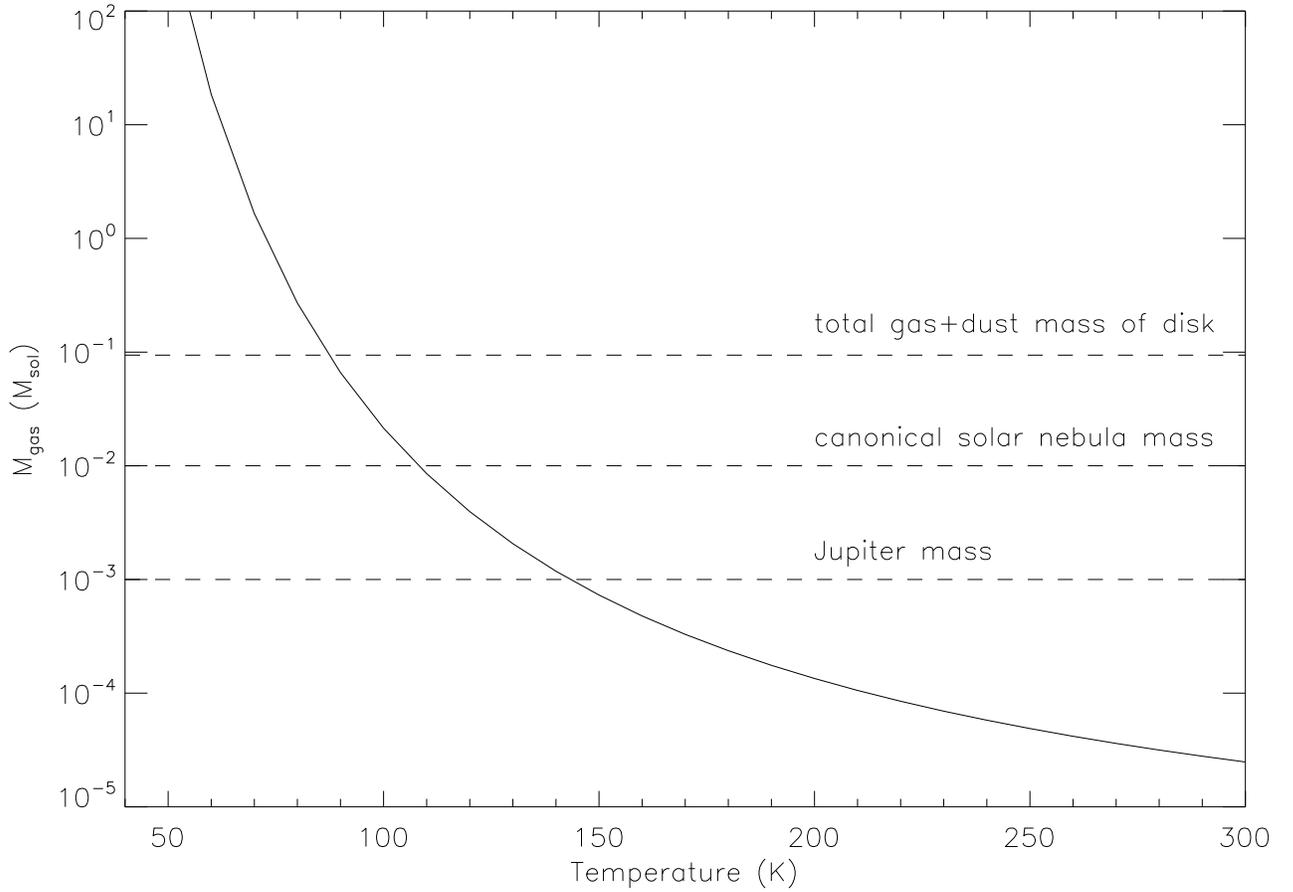}
\caption{\label{H2gas} H$_{2}$ gas mass inferred from the strength of the 17 $\mu$m S(1) emission line as a function of temperature (solid curve). The horizontal dashed lines represent the masses of Jupiter, canonical solar nebula, and the total gas+dust mass of the disk around V4046 Sgr \citep{rosenfeld2013}.}
\end{figure}

Figure \ref{H2gas} shows the resulting mass of the emitting H$_{2}$ gas for various excitation temperatures, as well as the total gas+dust mass of the disk established by \citet{rosenfeld2013} (0.094 M$_{\odot}$). It is apparent that if the H$_2$ gas were at temperatures $\lesssim$100 K, the gas mass needed to produce the line flux we see in the {\it Spitzer} spectrum would exceed that of the total gas+dust disk mass. Indeed, from the upper limit on the flux of the H$_{2}$ S(0) emission feature at 28.2 $\mu$m ($<$1.9 $\times 10^{-14}$ erg s$^{-1}$ cm$^{-2}$), we estimate the lower limit for the excitation temperature of the H$_{2}$ emitting gas to be $\sim$100 K \citep[see] [eqn. 3]{thi2001}. This suggests that much of the IR-emitting H$_{2}$ gas resides in the surface layers of the disk, where the temperature is warmer than the midplane at a given radius due to stellar X-ray and/or UV irradiation.

\subsection{[Ne~{\sc ii}] and [Ne~{\sc iii}]}
The brightest line in the {\it Spitzer} IRS spectrum of V4046 Sgr is that of [Ne~{\sc ii}] at 12.8 $\mu$m. We measure a [Ne~{\sc ii}] emission flux of 8.67 $\pm$ 0.13 $\times 10^{-14}$ erg s$^{-1}$ cm$^{-2}$. [Ne~{\sc ii}] emission is often prominent in T Tauri systems and likely traces hot  (T $\sim$5000 K) disk gas that is being photoevaporated by EUV and X-ray radiation incident on the disk's upper atmosphere within a few tens of AU of the central star \citep{glassgold2004}. Irradiated disk models \citep{meijerink2008, ercolano2008, ercolano2010} predict a correlation between the strength of the [Ne~{\sc ii}]  line and the X-ray luminosity. Recent studies of circumstellar disks \citep[e.g.][]{sacco2012, gudel2010} have shown a weak correlation between X-ray luminosity and [Ne~{\sc ii}], and blueshifts of [Ne~{\sc ii}] consistent with emission due to a photoevaporative wind or magnetically driven outflow. 

\citet{sacco2012} obtained high resolution mid-IR spectra covering the [Ne~{\sc ii}] (12.81 $\mu$m) emission line towards V4046 Sgr (and 31 other YSOs) using the VISIR spectrograph on the VLT. They detected V4046 Sgr and measured a [Ne~{\sc ii}]  line flux of 6.6 $\pm$ 0.2 $\times$ 10$^{-14}$ erg  s $^{-1}$ cm$^{-2}$.  \citet{sacco2012} also measured the flux of the [Ne~{\sc ii}] 12.81 $\mu$m emission from the {\it Spitzer} spectrum as 7.68 $\pm$ 0.22 x 10$^{-14}$ erg  s $^{-1}$ cm$^{-2}$, which is consistent with our flux measurement after rescaling for slit loss. The difference between the {\it Spitzer} and VLT [Ne {\sc ii}] flux measurements  may be indicative of time variability of the emission, but could also be due to the different slit sizes. 

The VLT [Ne~{\sc ii}] emission from V4046 Sgr is slightly blueshifted and likely arises from a photoevaporative wind is escaping the inner gaseous disk \citep{sacco2012}. \citet{rapson2015} show that the R$\sim$30 AU gap in the disk \citep{rosenfeld2013} is filled with (sub)-micron dust grains that may provide enough opacity to produce the observed [Ne {\sc ii}] line. Calculation of the 12 $\mu$m opacity throughout the V4046 Sgr disk based on dust column densities from \citet{rosenfeld2013} suggests that there is sufficient optical depth at R$\sim$ 20-50 AU to absorb the red-shifted portion of the [Ne II] emission. Thus, the [Ne II] emission likely originates in this region of the inner disk, as \citet{sacco2012} suggested.

\citet{espaillat2007} measure the [Ne {\sc ii}] emission from CS Cha, a $\sim$2 Myr old T Tauri star that is similar to V4046 Sgr in that is surrounded by a transitional disk with an inner gap out to R$\sim$43 AU. V4046 Sgr has a [Ne {\sc ii}] luminosity of $5.5 \times 10^{28}$ erg s$^{-1}$ (assuming D=73 pc) which is roughly half that measured for CS Cha. If the strength of the [Ne {\sc ii}] emission is due to X-ray and EUV photons impinging on the disk, then this lower [Ne {\sc ii}] luminosity would be consistent with the fact that the X-ray luminosity of V4046 Sgr is also lower than that for CS Cha \citep{donati2011,feigelson1993}. However, both disks have a [Ne {\sc ii}] luminosity higher than what is typically observed for classical T Tauri stars \citep[][and references therein]{espaillat2007, pascucci2007}. \citet{espaillat2007} attribute this to the high accretion rate of CS Cha (1.2$\times$10$^{-8}$~M$_{\odot}$ yr$^{-1}$), yet V4046 Sgr has a much lower accretion rate \citep[5$\times$10$^{-10}$~M$_{\odot}$ yr$^{-1}$;][]{donati2011} and still shows a high [Ne {\sc ii}] luminosity. Therefore, mass accretion rate may not be a good indication of the strength of the [Ne {\sc ii}] line in circumstellar disks.  This comparison points out the need for additional studies aimed at investigating potential correlations between [Ne {\sc ii}] line luminosity and star/disk system parameters.

The 12.8 $\mu$m [Ne~{\sc ii}] feature, combined with measurements of emission from [Ne~{\sc iii}], can be used as X-ray/EUV radiation field diagnostics. Emission from 15.6 $\mu$m [Ne~{\sc iii}] is rarely detected in T Tauri systems, and upper limits are usually an order of magnitude below that of [Ne~{\sc ii}] \citep[e.g.,][]{lahuis2007}. We report only an upper limit for the 15.6 $\mu$m [Ne~{\sc iii}] line in Table \ref{lines}, as this line is at best tentatively detected and a potential line is only apparent in one nod. For V4046 Sgr, we find [Ne~{\sc iii}]/[Ne~{\sc ii}] $\lesssim$ 0.13, which is consistent with the ratio determined for TW Hya ([Ne~{\sc iii}]/[Ne~{\sc ii}] $\sim$ 0.045) \citep{najita2010}, and is comparable to other CTTS that lack jets \citep[e.g.][]{lahuis2007}. Modeling of X-ray irradiation of protoplanetary disks \citep{ercolano2008} also predicts [Ne~{\sc iii}]/[Ne~{\sc ii}] $\lesssim$ 0.1 for X-ray bright systems.

\citet{pascucci2014} constrain the EUV irradiation of the V4046 Sgr disk by studying radio data that traces free-free emission from ionized gas caused by EUV irradiation. They find that the inferred EUV luminosity reaching the disk is much lower for older disks (V4046 Sgr, TW Hya and MP Mus) than for younger, less evolved disks, and that the EUV flux for V4046 Sgr is not sufficient to produce the [Ne {\sc ii}] luminosities measured by \citet{sacco2012} and confirmed in this work. Thus, the disk is likely being photoevaporated by $\sim$1 keV X-ray photons at a faster rate than if EUV photons were responsible for the photoevaporation.

\subsection{H$_{2}$O and OH}

Emission from H$_{2}$O and OH has been detected in the Spitzer spectra around many young stars \citep[e.g.,][]{carr2011,pontoppidan2010,salyk2008,carr2008}. In general, this emission is found to arise from regions of the disk with temperatures $\lesssim$1000 K, corresponding to a radius of a few AU, where some H$_{2}$O is photodissociated into OH by far-UV photons. V4046 Sgr shows tentative evidence for H$_{2}$O emission in the {\it Spitzer} spectrum, but lacks emission at wavelengths that typically trace the presence of H$_{2}$O and OH in circumstellar disks (see section 3.2.1). In particular, the Spitzer spectrum of TW Hya \citet{najita2010} shows no emission from H$_{2}$O in the {\it Spitzer} SH spectrum, but displays many OH emission lines. 

Emission from the higher energy rotational states (E$_{upper} \gtrsim$7000 K) of OH observed in the TW Hya spectrum likely arises from warm (T$\sim$1000 K) H$_{2}$O gas that has been photodissociated within the inner 1 AU of the disk. This suggests that, even though TW Hya exhibits an inner hole (R$\sim$4 AU) in its dusty disk \citep{hughes2007, rosenfeld2012a}, there is gas present within this hole in the form of dissociated water vapor. 
We cannot rule out the presence of OH emission in the V4046 Sgr IRS spectrum at the level of that in the TW Hya IRS spectrum, because the V4046 Sgr spectrum is less sensitive (due to shorter exposure times\footnote{V4046 Sgr was observed with {\it Spitzer} for 30 and 14 seconds in the SH and LH mode, respectively, while TW Hya was observed, only in SH mode, for 120 and 600 seconds.}). Thus, while the lack of H$_{2}$O emission in the {\it Spitzer} spectrum of V4046 Sgr may be due to FUV radiation photodissociating the H$_{2}$O to form OH, more sensitive measurements and modeling of emission features are needed to confirm the presence of OH.

Emission from OH is clearly present in the {\it Herschel} spectra of V4046 Sgr, while o-H$_{2}$O $8_{18} \rightarrow 7_{07}$ emission at 63.32 $\mu$m (Figure \ref{herschel_zoom}) is tentatively detected. \citet{howard2013}, \citet{riviere_marichalar2012b}, and \citet{keane2014} detect ortho-H$_{2}$O  $8_{18} \rightarrow 7_{07}$ emission from various T Tauri disks in the Taurus/Auriga and other nearby star forming regions, especially from those that also display [O~{\sc i}] emission. Based on visual inspection of the {\it Herschel} spectra in \citet{thi2010}, TW Hya may also show o-H$_{2}$O emission at 63$\mu$m, though it is not noted in that paper. Emission from 63 $\mu$m o-H$_{2}$O in T Tauri disks had been thought to originate from the same regions at which the hot H$_{2}$O gas emission in {\it Spitzer} spectra is present \citep{pontoppidan2009, meijerink2009, salyk2008}, but recent modeling of o-H$_{2}$O emitting sources  \citep{riviere_marichalar2012b} has shown that the 63 $\mu$m emission originates at larger radii in the disk. Similarly detailed modeling of the emission spectrum of V4046 Sgr is needed to confirm if this is the case for V4046 Sgr; if so, it is a direct probe of cooler H$_{2}$O in the disk.

\citet{hogerheijde2011} detect o-H$_{2}$O $1_{10} \rightarrow 1_{01}$ and p-H$_{2}$O $1_{11} \rightarrow 0_{00}$ emission from TW Hya at 538 and 269 $\mu$m, respectively, with {\it Herschel}-HIFI. They attribute this emission to a large reservoir of water vapor at r $>$ 50 AU from the central star. We do not detect emission from these H$_{2}$O transitions in our (relatively low sensitivity) SPIRE spectrum.
 
We also detect collisionally excited OH $^{2}\Pi_{3/2}$ emission towards V4046 Sgr at 119.23 and 119.44 $\mu$m in the PACS range spectra, along with radiatively excited OH $^{2}\Pi_{3/2}$ emission at 84.60 and 84.42 $\mu$m (blended with CO) in the PACS line spectra (Figure \ref{herschel_zoom} and \ref{herschel_line}, respectively). \citet{fedele2012, fedele2013} and \citet{wampfler2013,wampfler2010} detect H$_{2}$O and OH emission from protoplanetary disks around both low and high-mass young stellar objects. Emission from both the collisionally excited  $^{2}\Pi_{3/2}$ level, as well as the far-infrared radiatively excited OH $^{2}\Pi_{1/2}$ level, is seen in most systems. \citet{fedele2013} also detect OH emission features towards protoplanetary disks around both low and high mass stars. Since emission from OH $^{2}\Pi_{1/2}$ transitions is not detected in the PACS range scan spectra of V4046 Sgr, we conclude that  collisionally excited OH molecules may dominate emission around more evolved sources like V4046 Sgr, whereas far-infrared pumping and collisional excitations are both important processes in younger, less evolved systems. 

\subsection{[O~{\sc i}]}
Emission from [O~{\sc i}] is well detected at 63.18 $\mu$m in the {\it Herschel}/PACS spectrum of V4046 Sgr (Figure \ref{herschel_zoom}, Table \ref{lines_herschel}). [O~{\sc i}] emission arises in protoplanetary disks when UV radiation from the central T Tauri star photodissociates CO or OH molecules in the outer disk \citep{mamon1988,aresu2012}. Disk gas mass estimates and other disk properties can be derived through observations of $^{12}$CO and $^{13}$CO emission (especially when these features are optically thin) so it is important to understand how the CO in a disk has been affected by such a UV field. Emission from [O~{\sc i}] has been detected around both T Tauri and Herbig Ae/Be stars \citep[e.g][]{keane2014, fedele2013, howard2013, riviere_marichalar2012a, riviere_marichalar2012b, mathews2010}, and is also well detected in the {\it Herschel} spectrum of TW Hya  \citep{thi2010}. The [O~{\sc i}] luminosity at 63 $\mu$m from TW Hya is comparable to the emission we measure from V4046 Sgr, yet neither V4046 Sgr nor TW Hya show detectable 145 $\mu$m [O~{\sc i}] emission or 158 $\mu$m [C {\sc ii}] emission \citep{thi2010}. Emission from 370.3 $\mu$m [C {\sc i}] is also not detected in the V4046 Sgr, but a 3$\sigma$ upper limit is reported in Table \ref{lines_herschel}. This may suggest that UV radiation from the central stars in both systems is actively photodissociating CO and/or OH in the outer layers of the disk \citep{acke2005}, with the excess C and/or H then being bound up in the form of carbon or hydrogen-bearing molecules.

\citet{keane2014} used Herschel PACS to search for  [O~{\sc i}] 63 $\mu$m emission from full disks (i.e. systems with no evidence for a disk gap), transition disks, and outflow systems in star forming regions. They detect [O~{\sc i}] emission from 21 transitional disks and find that the strength of the [O~{\sc i}] lines is typically $\sim$1/2 that of full disk sources. Of the stars in the \citet{keane2014} sample, V4046 Sgr is most similar to the full disk systems DK Tau, DQ Tau, DS Tau, DG Tau, and HK Tau in Taurus in terms of spectral type, multiplicity, and accretion rate. Assuming a distance to Taurus of 137 pc \citep{torres2008}, we find that the luminosity of the [O~{\sc i}] line for V4046 Sgr is $\sim$1/4 that of these Taurus star/disk systems, consistent with the notion that transitional disk systems tend to have lower [O~{\sc i}] line strengths \citep{keane2014}.

\subsection{HCN}

HCN serves as a tracer of carbon-rich disk regions, where there is an excess of carbon to bond with free H and N atoms. Normally, most of the carbon is bound in either CO or CO$_{2}$ molecules, but in disks with a high H$_{2}$O content, much of the oxygen is locked up in H$_{2}$O, leaving behind excess C \citep{agundez2008}.  Emission from mid-IR HCN is commonly seen in T Tauri stars \citep{pontoppidan2009} and is potentially correlated with accretion rate and X-ray luminosity \citep{teske2011}.  We detect emission from HCN at $\sim$14 $\mu$m in the {\it Spitzer} spectrum of V4046 Sgr (Figure \ref{spitzer_spec}). 
The integrated HCN line flux is comparable to HCN emission from classical T Tauri stars in Taurus \citep[3.2 mJy $\mu$m;][]{teske2011}. HCN has previously been detected around V4046 Sgr at submm wavelengths with the IRAM 30m and APEX telescopes \citep{kastner2008, kastner2014} and this submm emission may also be correlated with UV/X-ray luminosity \citep{kastner2008}. 
 
\subsection{CO}

CO emission from the outer (R $\gtrsim$ 30 AU to $\sim$ 370 AU) disk of V4046 Sgr has previously been detected in the submm with the SMA \citep{rodriguez2010} and IRAM \citep{kastner2008}, and was extensively modeled by \citet{rosenfeld2013}. The {\it Herschel} PACS data extend these CO detections to higher J transitions that probe the warmer inner regions of the disk (R $\sim$10-30 AU). The PACS line scan data (Figure \ref{herschel_line}) show emission from high-J transitions of CO  that are not apparent in the less sensitive range scan PACS data (Figure \ref{herschel_spec}). Specifically, we detect emission from CO J= 31$\rightarrow$30 (blended with OH), 18$\rightarrow$17, and 15$\rightarrow$14, and tentatively detect emission from CO J=30$\rightarrow$29. Modeling of these {\it Herschel}/PACS CO emission features, along with the other emission features mentioned above and previous studies of CO emission, would help constrain the gas temperature distribution throughout the V4046 Sgr disk. 

\section{Silicate Dust}
\subsection{Dust grain modeling}

\begin{figure}
\includegraphics[scale=1]{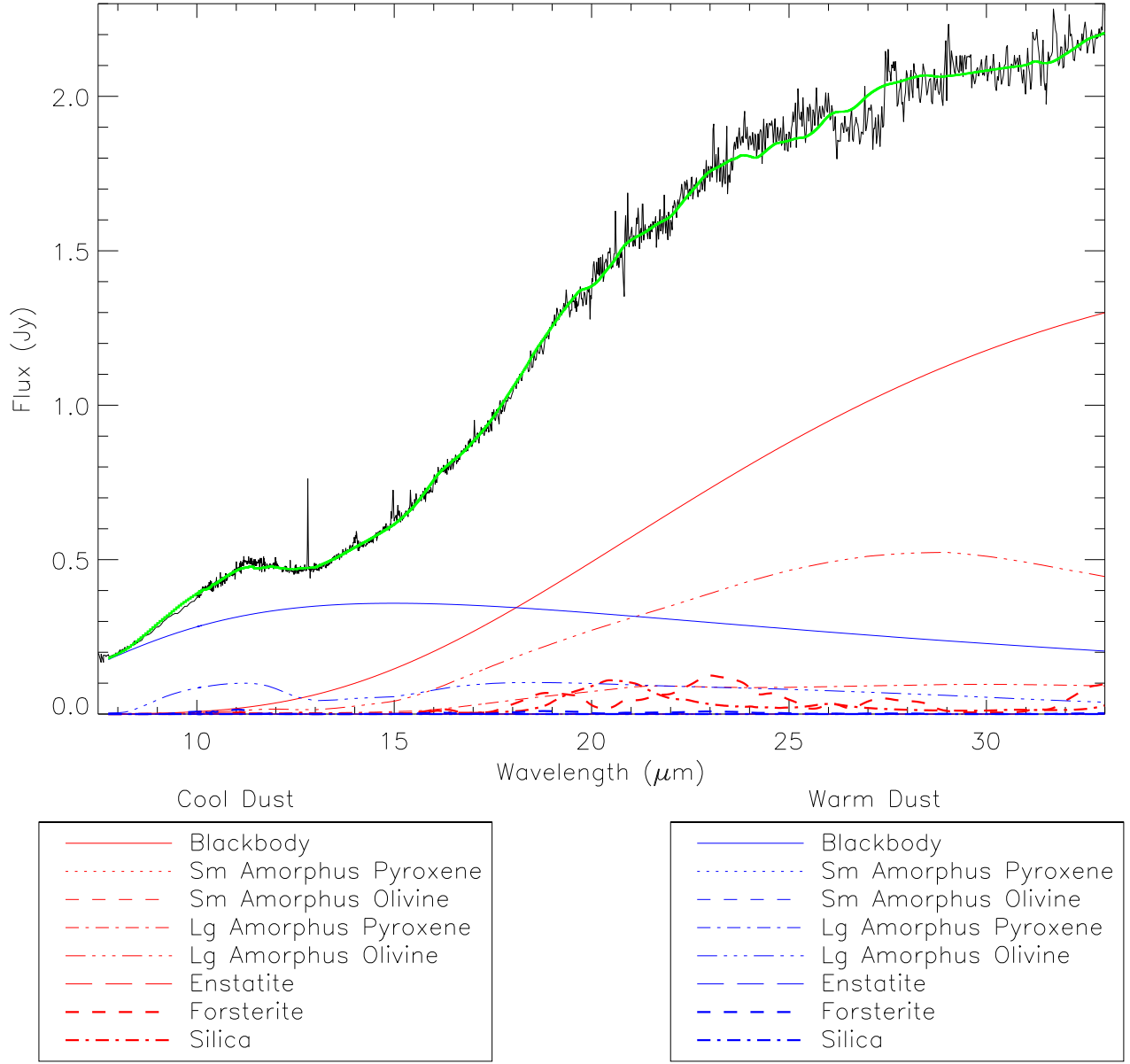}
\caption{\label{silicate_model} {\it Spitzer} IRS spectra (black) with best fit silicate model overlaid (green). The cool dust components of the model are shown in red, and the warm dust components are shown in blue.}
\end{figure}

Silicate grain spectral emission features, including amorphous silicates, as well as crystalline grain species such as forsterite, enstatite, silica, pyroxene and olivine,  have been identified in the mid-IR spectra of many T Tauri stars \citep[e.g.][]{cohen1985, honda2003, sargent2006, sargent2009_2}. Analysis of such spectra yields dust grain compositions and, in particular, the type and relative concentrations of silicates in these disks. We modeled the silicate dust features in the {\it Spitzer} IRS spectrum of V4046 Sgr using custom IDL programs developed by \citet{sargent2009_1}. In this method, one fits a two-temperature model that is the sum of optically thick emission from warm and cold blackbodies (T$_{w}$ and T$_{c}$, respectively), and optically thin emission from two dust populations corresponding to T$_{w}$ and T$_{c}$, respectively. The model flux is given by
\begin{equation}
\begin{array}{l}
\displaystyle F_{\nu}(\lambda)^{mod} = B_{\nu}(\lambda, T_{c}) \Big{[} \Omega_{c} +  \displaystyle \sum_{j} a_{c,j} \kappa_{j} (\lambda) \Big{]} +\\
\displaystyle \hspace{18mm} B_{\nu}(\lambda, T_{w}) \Big{[} \Omega_{w} +  \displaystyle \sum_{j} a_{w,j} \kappa_{j} (\lambda) \Big{]} ,
 \end{array}
\end{equation}
where B$_{\nu}(\lambda,T)$ is the Planck function, $\Omega_{c}$ ($\Omega_{w}$) is the solid angle of the blackbody at temperature T$_{c}$ (T$_{w}$), a$_{c,j}$ (a$_{w,j}$) is the mass weight of silicate dust feature \emph{j} at temperature T$_{c}$ (T$_{w}$), and $\kappa_{j}(\lambda)$ is the opacity at wavelength $\lambda$ for dust species \emph{j}. While a multi-temperature model would be more realistic,  a two-temperature dust model that characterizes the inner and outer disk regions generally well describes the thermal IR emission from T Tauri star disks \citep{sargent2009_2}.
To determine the best fit model to the V4046 Sgr spectrum, we minimize $\chi^{2}$ between the model and the 7.7 - 33 $\mu$m region of the spectrum. It is assumed, for simplicity, that all data points are independent (i.e., the covariance between pixels is zero). Our {\it Spitzer} data is excessively noisy longward of 33 $\mu$m, so we omit these data from the fit. The best fit model (Table \ref{silicate_comp}), which has 18 free parameters fitting 1490 individual data points, is shown in Figure \ref{silicate_model}. This model spectrum assumes a central source of blackbody temperature 4200 K, and indicates that the temperature of dust detected by {\it Spitzer} lies in the range $\sim$120-340 K. 

\subsection{Dust Composition}
Table \ref{silicate_comp} lists the main results of the model fitting just described, in the form of the percent by mass of warm and cool dust grain species responsible for the mid-IR emitting region of the disk around V4046 Sgr. Note that this model does not include carbon or other featureless dust grains that contribute to the continuum emission. The modeling indicates that the silicate dust in the inner disk is primarily (86\% by mass) composed of large ($\sim$5 $\mu$m radius) amorphous pyroxene and olivine grains at both warm and cool temperatures. There are also clear signatures of warm and cool forsterite and silica in crystalline form; these grain types constitute the other 14\% of the total silicate dust composition, according to the model fitting. The modeling indicates that contributions to the mid-IR emission from small grains below the Rayleigh limit (2$\pi a/ \lambda \ll$ 1, where $a$ is the radius of the dust grain) or other non-crystalline components are negligible.

The foregoing results allow us to compare the composition of the dust in the circumbinary disk orbiting V4046 Sgr with other protoplanetary disks and other transitional disk objects of similar age, such as TW Hya and Hen 3-600 (TWA 3). Modeling of the {\it Spitzer} IRS spectra of TW Hya implies a large silicate grain fraction of $\sim$34.4\%, with only 2.5\% crystallinity, while modeling of Hen 3-600A results in a similar large silicate grain percentage of 31.9\%, but with a high crystallinity of 36.2\% \citep{sargent2006, honda2003}. Studies of T Tauri stars in the Taurus/Auriga star forming region and the TW Hydrae association have shown that these transitional disk systems have negligible crystallinity and only modest  large silicate grain fractions \citep[$<$35\%;][]{sargent2006}, yet the disk around V4046 Sgr has a comparatively high crystallinity fraction and is otherwise dominated by large silicate grains. 

\citet{sargent2009_2} use methods identical to those described above to model the {\it Spitzer} spectrum of 64 protoplanetary disks in the Taurus/Auriga star forming region \citep[age $\sim$1-2 Myr;][]{kenyon1995}, which is considerably younger than V4046 Sgr. They find median mass fractions for warm and cool crystalline silicate dust of 11\% and 15\%, respectively, whereas V4046 Sgr displays 3\% and 14\% crystallinity, respectively. However, the transitional and pre-transitional sources in the \citet{sargent2009_2} sample all have negligible crystallinity fractions. The fraction of large, cool dust grains inferred for the V4046 Sgr disk is also very high-- larger than 93\% of the sources in Taurus/Auriga. This suggests that grain growth has occurred in the disk, which is consistent with the advanced age of the system. \citet{sargent2009_2} also find that disks orbiting multiple star systems tend to have larger warm large grain fractions than single star systems. This may be due to dynamical interactions between the stars and the disk that trigger grain growth in the inner disk and/or to photoevaporation of small grains. Since V4046 Sgr is a multiple star system with a large mass fraction of cool large grains rather than warm large grains, it is possible that a planetary companion in the inner disk is clearing away smaller particles and enhancing the large grain population at larger (cooler) radii.

A variety of mechanisms have been suggested as to how crystalline silicates form in protoplanetary disks \citep[see, e.g.,][]{sargent2006}. One of the leading mechanisms involves amorphous silicates being thermally annealed into crystalline silicates as a result of heating induced by shock fronts within the disk. These shock fronts can result from star-disk interactions, if the disk is massive enough, or from local gravitational instabilities associated with planet formation. \citet{rapson2015} present direct-imaging evidence for dust segregation by size, as well as dust ring structure, in the inner $\sim$40 AU of the disk around V4046 Sgr, and suggest that planet formation has occurred or is ongoing in the disk. This planet formation activity, if present, would help explain the combination of moderate crystallinity and large dust grains. Grain growth and planet formation may have occurred recently in the disk, thus creating a larger than typical crystallinity fraction for a disk with an inner gap and advanced age. Crystalline silicates at or interior to these planet formation zones would likely accrete onto either the central binary star or protoplanet(s). Since the {\it Spitzer} spectra are tracing the inner few AU of the disk (see section 4.8) we may be seeing the crystalline silicates in the innermost portion of the disk that have not yet accreted onto the central stars. 

Another possible scenario for the presence of crystalline silicates in the disk, and the varying crystallinity fraction amongst young stars, is thermal annealing of surface grains due to intense pre-MS stellar outbursts \citep{abraham2009}. Such outbursts are typically associated with  episodes of dramatically enhanced pre-MS accretion rate, however, and the (highly evolved, low-accretion-rate) V4046 Sgr system is unlikely to have experienced such an episode. Furthermore, our modeling shows that the warm dust is currently at a temperature of $\sim$350 K, far below the  $\sim$700 K required to thermally anneal dust, so if crystalline silicates formed via irradiation during an accretion-related outburst, they have since cooled and migrated outward in the disk.

\subsection{Amending the \citet{rosenfeld2013} model }

Modeling of the entire V4046 Sgr disk was conducted by \citet{rosenfeld2013} based on their SMA CO data, as well as on {\it Spitzer} IRS spectra and archival photometric data of V4046 Sgr. Using a 3D radiative transfer code, they developed a three-component model that includes an inner ``gap" filled with $\mu$m-sized dust grains out to $\sim$ 30 AU, a ring of cm/mm-sized dust grains from $\sim$30-45 AU,  and an extended halo of CO and small grain emission. Their resulting model well reproduces the mid- to far-IR photometry and 1.3 mm $^{12}$CO and continuum interferometric observations of V4046 Sgr. The same model fits our {\it Herschel} data remarkably well, due to their inclusion of $\mu$m-sized grains out to $\sim$ 30 AU. However, the \citet{rosenfeld2013} model did not reproduce the detailed SED structure at {\it Spitzer} IRS wavelengths. The mismatch between the model and data in this wavelength range was due to the presence of a complex (crystalline plus amorphous) silicate grain mixture in the inner disk (see section 4.7) that is not accounted for by the three-component model. 

\begin{figure}
\includegraphics[scale=1]{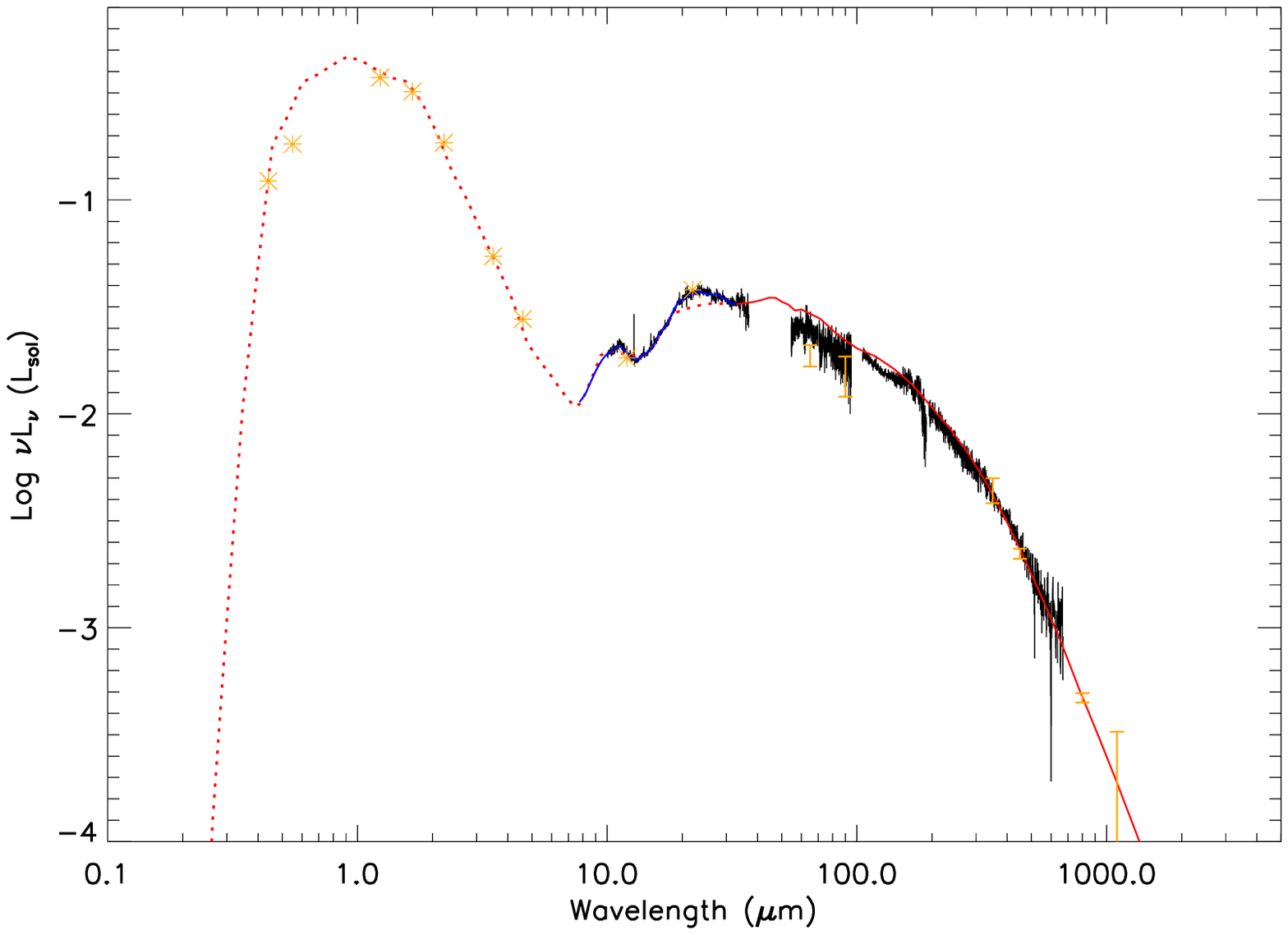}
\caption{\label{new_fit} {\it Spitzer} and {\it Herschel} spectra (black) with our silicate dust model (blue) from 7-33 $\mu$m and the \citet{rosenfeld2013} model (red solid) from 33-600 $\mu$m overlaid. The \citet{rosenfeld2013} model from 0.25-33 $\mu$m is shown as the red dotted line. Orange asterisks are photometric data points from the literature (B \& V from \citet{hog2000}, I from \citet{messina2010}, J,H \& K from \citet{cutri2003}, AKARI 65 \& 90 $\mu$m data from \citet{yamamura2010}, WISE bands 1-4 data from  \citet{cutri2013}, and 350,450, 800 \& 1100 $\mu$m data from \citet{jensen1996}.}
\end{figure}

\begin{figure}
\includegraphics[scale=1]{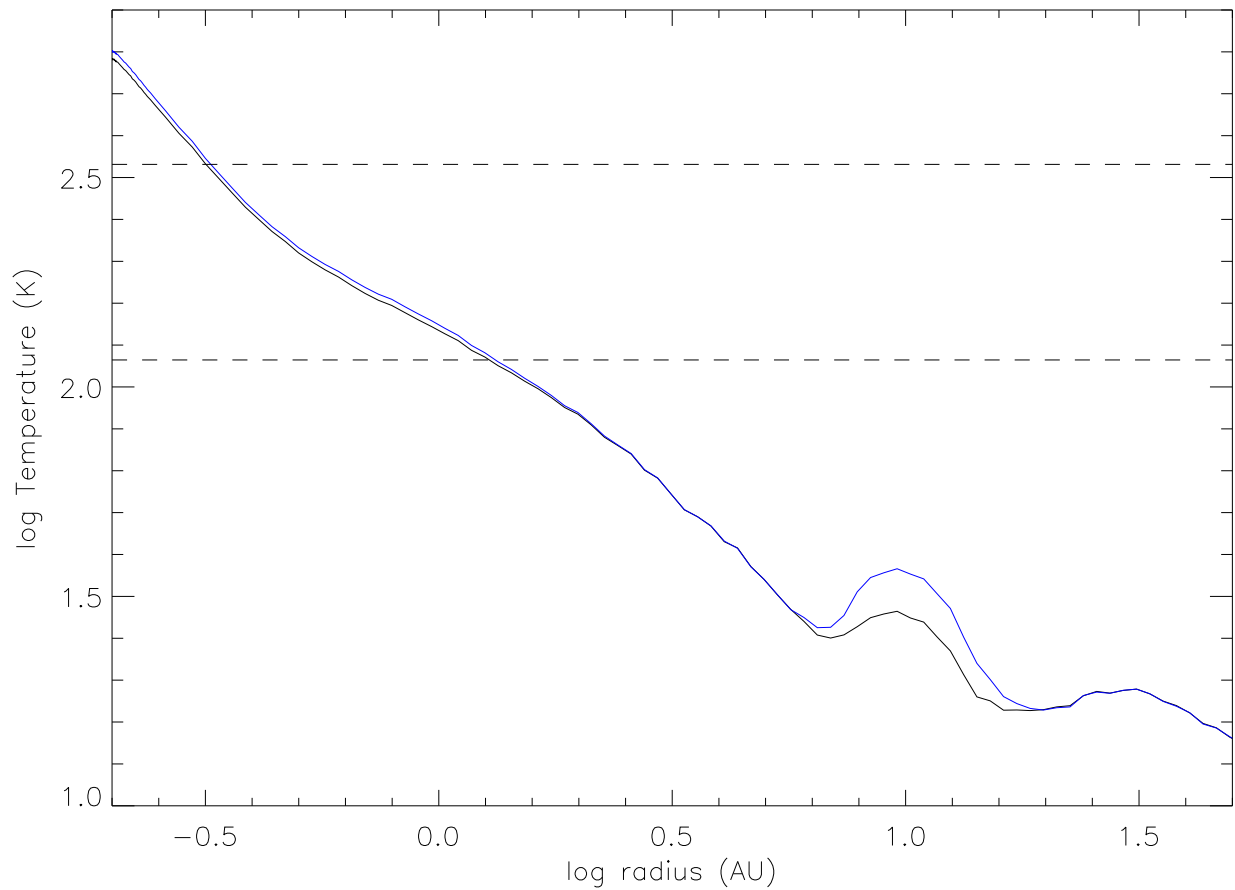}
\caption{\label{temp_rad} Temperature at a given disk radius in the \citet{rosenfeld2013} model for large grains (R $>$5 $\mu$m; black solid line) and small grains (R $\sim$5 $\mu$m; blue solid line). The horizontal dashed lines at 340 K and 116 K represent the two temperatures of the micron-sized silicate dust grains in our model.}
\end{figure}

In Figure \ref{new_fit}, we replace the shorter-wavelength (7.7-33 $\mu$m) portion of the \citet{rosenfeld2013} model with our silicate model. This revised model, which better reproduces the {\it Spitzer} spectra, merges seamlessly with the Rosenfeld model at $\sim$33 $\mu$m. Using the \citet{rosenfeld2013} model, we can estimate the characteristic disk radii probed by our silicate model, assuming that our newly calculated dust composition and resulting dust opacities do not significantly alter the disk temperature in their models. Figure \ref{temp_rad} shows disk temperature versus radius for both large (mm-sized) and small (micron-sized) grains based on the \citet{rosenfeld2013} model. Our silicate dust modeling shows dust at 340 K and 116 K, which correspond to a radius of 0.3 AU and 1.3 AU, respectively, for small ($\mu$m-sized) grains. We conclude that the {\it Spitzer} data likely probe the dust grain populations of the inner disk within these approximate radii.

\section{Conclusions}

We have presented an analysis of {\it Spitzer} and {\it Herschel} spectra of the disk around V4046 Sgr that elucidates the gas and dust constituents within the disk. We confirm the measurement of strong [Ne~{\sc ii}] at 12.8 $\mu$m previously reported by \citet{sacco2012} and report an upper limit for [Ne~{\sc iii}] emission.
The [Ne~{\sc iii}]/[Ne~{\sc ii}] ratio is consistent with other CTTS systems that are X-ray bright and lack jets. We also detect emission from pure rotational H$_{2}$ 0-0 S(1) (J=3$\rightarrow$1)  at 17 $\mu$m that likely originates in the surface layers of the disk where the temperature is warmer than the midplane at a given radius due to stellar UV/X-ray irradiation.

The X-ray/UV radiation coming from the central binary is likely photodissociating CO, H$_{2}$O, and OH, resulting in emission from [O~{\sc i}] and OH. We detect emission from [O~{\sc i}] at 63 $\mu$m, but do not detect [C~{\sc i}] or [C~{\sc ii}] in the mid- to far-infrared. Thus, the excess C atoms may be forming HCN and other hydrocarbons throughout the disk, and/or the [O~{\sc i}] is arising from photodissociated OH in the disk. The strength of the [O~{\sc i}]  emission is lower than that for star-disk systems of similar spectral type and accretion properties surrounded by continuous disks, as was found by \citet{keane2014} and \citet{howard2013}.  V4046 Sgr exhibits OH $^{2}\Pi_{3/2}$ emission at 84.4, 84.6, and 119 $\mu$m, and possible weak H$_{2}$O emission features in the mid- to far-infrared. This suggests that H$_{2}$O and collisionally excited OH molecules reside in the the outer layers of the V4046 Sgr disk. 

Modeling of the {\it Spitzer} spectra reveal that the mid-IR emitting region of the disk consists primarily of large ($\sim 5 \mu$m) silicate dust particles (86\% by mass), with crystalline forsterite and silica particles making up the other 14\% of the mass. Overall, the presence of large grains suggests that grain growth may be occurring in the inner $\sim$30 AU gap around V4046 Sgr and the abundance of cool dust suggests that dust may be settling towards the midplane where temperatures are lower at a given radius. The moderate crystallinity fraction is similar to that of less evolved protoplanetary disks in the Taurus/Auriga star forming region \citep{sargent2009_2}, and may result from heating and thermal annealing of micron-sized dust that is associated with planet formation activity in the inner disk \citep{rapson2015}. Combining our silicate dust model with the mid-IR to sub-mm three-component model of \citet{rosenfeld2013}, we find that the dust emission seen in the Spitzer spectrum likely originates from regions of the disk interior to 1.3 AU. These results together suggest that grain growth might, in part, explain the transitional appearance of this disk.

\section*{Acknowledgements}
\acknowledgments
We would like to thank Katrina Exter for assistance with processing the PACS and SPIRE data, as well as Wing Fai Thi for his helpful discussions. We would also like to thank the anonymous referee for their thoughtful comments and suggestions. This research work is based in part on observations made with {\it Herschel}, a European Space Agency Cornerstone Mission with significant participation by NASA; support was provided by NASA through an award issued by JPL/Caltech. Additional support is provided by National Science Foundation grant AST-1108950 to RIT.

\bibliographystyle{apj}


\begin{deluxetable}{lccccc}
\tabletypesize{\small}
\tablecaption{Emission Lines in {\it Spitzer} IRS Spectrum of V4046 Sgr}
\tablewidth{0pt}
\tablehead{Species &  Wavelength\tablenotemark{a} & E$_{u}$ &Flux &  FWHM &   EW \\ 
 & ($\mu$m)  & K & ($10^{-14}$ erg s$^{-1}$ cm$^{-2}$) &  ($10^{-2} \mu$m) & ($10^{-3} \mu$m) }
\startdata

H {\sc i} 7-6		&	12.37	&	154854	&	 0.61 $\pm {0.17}$ 	&	0.32 $\pm {0.09}$ 	&	0.65 $\pm {0.18}$ \\ 
Ne~{\sc ii}			&	12.81	&	1124		&	8.67 $\pm {0.13}$ 	&	 1.51 $\pm{0.02}$  	&	10.00 $\pm{0.15}$\\ 
HCN	 			&	14.03	&	...		&	4.90 $\pm{0.38}$	&	8.00 $\pm{0.61}$ 	&	5.90 $\pm{0.45}$ \\
CO$_{2}$ 		&	14.97	&	...		&	 4.31 $\pm{0.25}$	&	 2.77 $\pm{0.16}$  	&	 5.06 $\pm{0.30}$\\ 
H$_{2}$O 16 -15$^{*}$ 	&	15.41	&	3602		&	1.49 $\pm{0.22}$ 	&	1.80 $\pm{0.27}$ 	&	 1.76 $\pm{0.26}$\\ 
Ne~{\sc iii}$^{*}$   	&	15.55	&	927		&	$<$ 1.11 	&	 ...	&	 ... \\ 
H$_{2}$ S(1)		&	17.03	&	1015		&	1.50 $\pm{0.24}$ 	&	 2.00 $\pm{0.33}$  	&	 1.62 $\pm{0.27}$\\ 
H$_{2}$O 9 - 8$^{*}$  &	23.86	&	1397		&	  1.79 $\pm{0.51}$	&	  2.74 $\pm{0.79}$	&	 1.80 $\pm{0.52}$\\

\enddata
\label{lines}
\tablenotetext{*}{~Tentative detection.}
\tablenotetext{a}{~Observed wavelength.}
\end{deluxetable}

\begin{deluxetable}{lccccc}
\tabletypesize{\footnotesize}
\tablecaption{Emission Lines in all {\it Herschel} Spectra of V4046 Sgr}
\tablewidth{0pt}
\tablehead{Species &  Wavelength\tablenotemark{a} & E$_{u}$ &Flux &  FWHM &   EW \\ 
 & ($\mu$m)  &  K & ($10^{-14}$ erg s$^{-1}$ cm$^{-2}$) &  ($10^{-2} \mu$m) & ($10^{-3} \mu$m) }
\startdata

\cutinhead{PACS SED}
[O~{\sc i}]$^{3}$P$_{1} \rightarrow ^{3}$P$_{2}$			&	63.18	&	 228	&	 1.71 $\pm{0.43}$ 	&	 1.30 $\pm{0.33}$	&	 6.72$\pm{1.71}$ \\ 
o-H$_{2}$O 8 - 7$^*$						&	63.31	& 	1071	&	  1.09 $\pm {0.27}$ 	&	1.14 $\pm {0.28}$	&	 4.181 $\pm {1.03}$ \\ 
OH $^{2}\Pi_{3/2}$ J=$\frac{5}{2} - \frac{3}{2}$    			&	119.25	&	 121	&	 1.11 $\pm{0.13}$	&	 9.05 $\pm{0.91}$	&	 12.74 $\pm{1.28}$ \\
OH $^{2}\Pi_{3/2}$ J=$\frac{5}{2} - \frac{3}{2}$   			&	119.43	&	 121	&	1.10 $\pm{0.12}$	&	 10.92 $\pm{1.20}$	&	 12.65 $\pm{1.40}$ \\
CO  J=17-16$^*$										&	153.28	&	846  &	1.38$\pm {0.25}$	&	13.23 $\pm{2.36}$	&	24.05$\pm {4.29}$\\
\cutinhead{PACS Line-scan}

 OH $^{2}\Pi_{3/2}$ J=$\frac{7}{2} - \frac{5}{2}$  $+$ CO J=31-30	&	84.42	&	 ...	&	1.03 $\pm{0.05}$ 	&	3.30$\pm{0.15}$ &	6.49 $\pm{0.29}$ \\
OH $^{2}\Pi_{3/2}$ J=$\frac{7}{2} - \frac{5}{2}$   	&	84.60 		&	 291	&	1.69 $\pm {0.05}$ 	&	3.29 $\pm{0.09}$  	&	 11.00 $\pm{0.30}$\\ 
CO J=30-29$^*$	&	87.18	&	 2565	&	0.72 $\pm{0.06}$	&	3.23 $\pm{0.26}$ 	&	5.21 $\pm{0.41}$ \\  
CO J=18-17	&	144.80	&	 945	&	0.52 $\pm{0.22}$ 	&	8.78 $\pm{0.38}$ 	&	 8.02 $\pm{0.34}$\\ 
CO J=15-14  	&	173.64	&	 664	&	0.84 $\pm{0.02}$ 	&	10.98 $\pm{0.28}$  	&	17.12 $\pm{0.44}$ \\ 
o-H$_{2}$O  2$_{1,2}$ - 1$_{0,1}$ +  CH$^{+}$  &	179.53	&	 114	&	0.23 $\pm{0.03}$ 	&	 6.72 $\pm{0.99}$  	&	 5.89 $\pm{0.87}$\\ 

\cutinhead{SPIRE}
C I			&	370.30	&	 63	&	 $<$ 0.64 	&	.... 	&	....
\enddata
\label{lines_herschel}
\tablenotetext{a}{~Observed wavelength.}
\tablenotetext{*}{~Tentative detection.}
\end{deluxetable}

\begin{deluxetable}{lc}
\tabletypesize{\normalsize}
\tablecaption{Model Dust Mass Percentages in the V4046 Sgr Disk}
\tablewidth{0pt}
\tablehead{Dust type & \% by mass}
\startdata

Cool Large Amorphous Pyroxene & 20.01 $\pm$ 5.79\% \\
Cool Large Amorphous Olivine & 65.93  $\pm$ 8.44\% \\
Cool Forsterite & 6.96  $\pm$ 3.24\%\\
Cool Silica & 7.09  $\pm$ 3.54\%\\
Warm Large Amorphous Olivine & 96.89  $\pm$ 15.55\%\\
Warm Forsterite & 3.11 $\pm$ 4.33\%\\
\hline
Cool Crystalline Silicates & 14.05 $\pm$ 5.83\%\\
Warm Crystalline Silicates & 3.11  $\pm$ 7.99\%\\
Total Large Silicates & 85.96  $\pm$ 10.20\%\\

\enddata
\label{silicate_comp}
\end{deluxetable}

\end{document}